\documentclass[fleqn,usenatbib]{mnras}

\usepackage[T1]{fontenc}
\usepackage{ae,aecompl}
\usepackage[dvipsnames]{xcolor}
\usepackage{ulem}

\usepackage{graphicx}	
\usepackage{amsmath}	
\usepackage{amssymb}	
\usepackage{tikz}
\usetikzlibrary{shapes,arrows}
\tikzstyle{block} = [rectangle, draw, fill=blue!20, 
text width=13em, text centered, minimum height=2em,font=\bfseries,line width=1pt]
\tikzstyle{line} = [draw, -latex', line width=3pt]
\tikzstyle{cloud} = [draw, ellipse,fill=red!20, node distance=3cm,
minimum height=4em]
\usepackage[flushleft]{threeparttable}
\usepackage[export]{adjustbox}
\usepackage{newtxtext,newtxmath}

\newcommand{\kms}{{\rm {km\, s^{-1}}}}
\newcommand{\msun}{{\rm M_{\odot}}}

\title[Retention and growth of IMBHs in NSCs]{Formation of supermassive black holes in galactic nuclei {\sc ii}: retention and growth of seed intermediate-mass black holes
}
\author[A. Askar et al.]{
Abbas Askar$^{1}$\thanks{E-mail: askar@astro.lu.se (AA)},
Melvyn B. Davies$^{2,1}$
and Ross P. Church$^{1}$
\\
$^{1}$Lund Observatory, Department of Astronomy, and Theoretical Physics, Lund University, Box 43, SE-221 00 Lund, Sweden\\
$^{2}$Centre for Mathematical Sciences, Lund University, Box 118, SE-221 00 Lund, Sweden
}

\date{Accepted XXX. Received YYY; in original form ZZZ}

\pubyear{2020}

\begin{document}
\label{firstpage}
\pagerange{\pageref{firstpage}--\pageref{lastpage}}
\maketitle

\begin{abstract}
In many galactic nuclei, a nuclear stellar cluster (NSC) co-exists with a supermassive black hole (SMBH). In this second in a series of papers, we further explore the idea that the NSC forms before the SMBH through the merger of several stellar clusters that may contain intermediate-mass black holes (IMBHs). These IMBHs can subsequently grow by mergers and accretion to form an SMBH. To check the observable consequences of this proposed SMBH seeding mechanism, we created an observationally motivated mock population of galaxies, in which NSCs are constructed by aggregating stellar clusters that may or may not contain IMBHs. Based on several assumptions, we model the growth of IMBHs in the NSCs through gravitational wave (GW) mergers with other IMBHs and gas accretion. In the case of GW mergers, the merged BH can either be retained or ejected depending on the GW recoil kick it receives. The likelihood of retaining the merged BH increases if we consider growth of IMBHs in the NSC through gas accretion. We find that nucleated lower-mass galaxies ($\rm M_{\star} \lesssim 10^{9}\,\msun$; e.g. M33) have an SMBH seed occupation fraction of about 0.3 to 0.5. This occupation fraction increases with galaxy stellar mass and for more massive galaxies ($\rm 10^{9}\,\msun \lesssim \rm M_{\star} \lesssim 10^{11}\,\msun$), it is between 0.5 and 0.8, depending on how BH growth is modelled. These occupation fractions are consistent with observational constraints. Furthermore, allowing for BH growth also allows us to reproduce the observed diversity in the mass range of SMBHs in the $\rm M_{\rm NSC} - M_{\rm BH}$ plane.

\end{abstract}


\section{Introduction}\label{sec:intro}

In many galactic nuclei, supermassive black holes (SMBHs) with masses between $10^{6} - 10^{10} \, \msun$) often co-exist with a nuclear stellar cluster (NSC). These are massive ($\sim 10^5$ to $10^{8}\,\msun$) and dense (up to $\rm 10^{7} \, \msun \, pc^{-3}$) stellar clusters which have typical radii of a few parsec. They are found in the nuclei of nearly 80 per cent of galaxies of all morphological types with stellar masses in the range $10^8$ to $10^{10}\,\msun$ \citep[see][and references therein]{Neumayer2020}. 

\begin{figure}
	\includegraphics[width=\columnwidth]{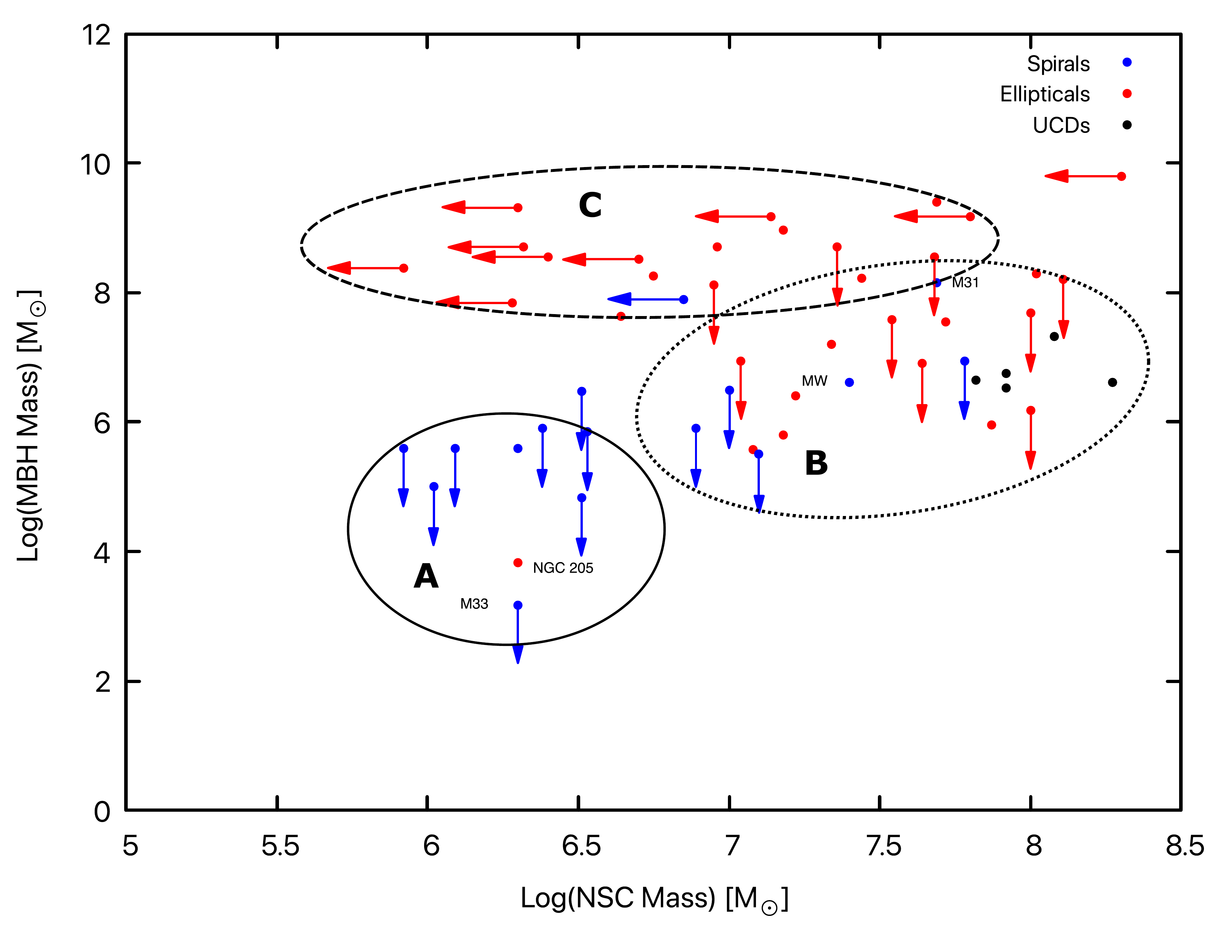}
	\caption{Figure shows observed galaxies for which \citet{Neumayer2020} (Table 3) have constraints on NSC Mass (x-axis) and SMBH Mass (y-axis). The coloured dots represent different morphologies for the host galaxy. The blue points are for spiral galaxies, the red points are for elliptical galaxies and the black are for ultra-compact dwarfs. We label three classifications of galaxies that are indicated by the circles. ``A'' Type (enclosed within solid black circle) galaxies are those for which NSC Masses are between $\sim \ 10^{6-6.5} \ \msun$ and have an upper limit on their BH mass between $10^{3-6} \ \msun$. This includes galaxies such as M33 and NGC 205. ``B'' Type (enclosed within dotted black circle) galaxies are those for which NSC Masses are between $\sim \ 10^{7-9.5} \ \msun$ and BH masses are between  $10^{6-8} \ \msun$. These include galaxies such as the Milky Way and M31. ``C'' Type (enclosed within dashed black circle) are elliptical galaxies with upper limits on NSC mass that are between $\sim \ 10^{5-7.5} \ \msun$ and have high BH masses larger than $10^{9} \ \msun$.}
    \label{fig:obs-nsc-vs-bh}
\end{figure}

There are about 50 observed galaxies for which there are observational constraints on the masses of NSCs and SMBHs \citep{Neumayer2020}. For a few of these galaxies, both the NSC and SMBH masses are well observed (e.g MW, M31) while for other galaxies upper limits for either BH mass or NSC mass are available. In Fig. \ref{fig:obs-nsc-vs-bh}, we show the NSC mass versus the SMBH Mass for these galaxies with data taken from \citet{Neumayer2020}. These observations reveal that low-mass galaxies with an NSC mass less than $\sim 3 \times 10^{6} \, \msun$ have either measurements or upper limits for BH mass that are between $10^{3}$ to $10^{6}$ $\msun$. This includes galaxies such as NGC 205 and M33 (categorized as ``A'' type galaxies in Fig. \ref{fig:obs-nsc-vs-bh}). For more massive spiral/S0 galaxies with NSC masses of $10^{7}$ to $10^{8} \, \msun$, the BH masses correlate with the NSC mass and are of the order of $10^{7}$ to $10^{8}  \, \msun$ (``B'' type galaxies in Fig. \ref{fig:obs-nsc-vs-bh}). These include galaxies such as Milky Way and M31. A significant fraction of elliptical galaxies contain massive BHs but have much lower NSC masses (``C'' type galaxies in \ref{fig:obs-nsc-vs-bh}). Masses of NSCs and SMBHs follow tight correlations with the properties of their host galaxies \citep{Ferrarese2006,leigh2015,graham2016proc,capuzzo2017} and this suggests that the formation and subsequent growth of both NSCs and SMBHs could be closely related \citep{antonini2015,Neumayer2020}. 

There are two main mechanisms for the formation and growth of an NSC: through the migration and merger of stellar clusters in the galactic nuclei due to dynamical friction \citep{tremaine1975,cd1993,oh2000,lotz2001,agarwal2011,Antonini2012,schiavi2021}, and through in-situ star formation from high density gas in the galactic nuclei \citep{loose1982,mihos1994,milosavljevi2004,aharon2015}. Observations of stellar populations within NSCs show that they have diverse ages and metallicities, which suggests that both cluster infall and in-situ star formation contribute to forming and growing the NSC \citep{Neumayer2020,2020arXiv200902335D,2020arXiv200902328A,bentley2021}. Based on a number of observational signatures, \citet{Neumayer2020} have suggested that there appears to be a transition in the NSC growth mechanism at galaxies of around $\rm M_{\star} \sim 10^{9} \msun$.
For instance, in galaxies with $\rm M_{\star} \lesssim 10^{9}\,\msun$ more than half of the stars in the NSC are metal-poor compared to the average metallicity for the host galaxy \citep{kacharov2018,fahrion2020,fahrion2021}. This lends credence to the idea that infall of low-metallicity globular clusters is the dominant contributor to the formation of the NSC. For galaxies with higher stellar masses and more massive NSCs, the metallicity of most stars in the NSC is higher than the typical metallicity of stars in that galaxy which suggests that they may have formed through the in-situ mechanism \citep{brown2018}. This suggests that the in-situ growth dominates in NSCs of massive galaxies. Additionally, at low masses, the scaling of NSC mass with galaxy stellar mass is in agreement with the globular stellar cluster inspiral scenario \citep{gnedin2014,sj2019, Neumayer2020}. Recent observations of low-mass galaxies in the Local Volume also suggest strong correspondence between globular clusters and NSCs \citep{carlsten2021,hoyer2021}. However at higher galaxy stellar masses, the scaling of NSC mass with galaxy mass is steeper, indicating that in-situ processes may contribute to the growth of the NSC in those galaxies \citep{Neumayer2020, pinna2021}. 

Many scenarios have been proposed to explain the formation and growth of SMBHs. Most of these scenarios suggest that SMBHs grew from smaller seed black holes (BHs), however, the exact seeding mechanism and the masses of the seed BHs are not known. These seed BHs could have originated from the evolution of massive stars or they may have formed with large initial masses of the order $10^{5} \, \msun$ through direct collapse of gas \citep{volonteri2010,johnson2013,greene2019rev,regan2020}. In the former case, BHs with masses in the intermediate-mass range of $10^{2}-10^{5} \, \msun$ ought to have existed in order to seed SMBHs. These intermediate-mass BHs (IMBHs) can also form through dynamical processes in dense stellar clusters such as globular clusters or young massive clusters \citep{Miller02,SPZ04,Giersz15,rizutto2020,gonzalez2021, dicarlo2021}.

\defcitealias{askar2021}{ADC2021}
In \citet[][abbreviated hereafter as \textcolor{blue}{ADC2021}]{askar2021}, we considered the idea that the NSC forms before the SMBH through the merger of several stellar clusters in a galactic centre \citep{ebisuzaki2001,kim2004,spz2006}. Some of these stellar clusters may deliver intermediate-mass black holes (IMBHs) to the galactic centre \citep{AMB2014,petts2017,sedda-gualandris2018}, where these IMBHs can subsequently grow and become SMBHs. In \citetalias{askar2021}, we carried out direct \textit{N}-body simulations of the merger of three stellar clusters to form an NSC to investigate the outcome of simulated runs containing zero, one, two and three IMBHs. If a single IMBH is delivered to the galactic nuclei, it can be retained there and can subsequently grow from gas accretion \citep{das2021,natarajan2021} or tidal capture and disruption of stars \citep{stone2017,alexander2017,boekholt2018,dittman2020}. If multiple IMBHs are delivered to the NSC, then we found that an IMBH binary is likely to form and merge due to gravitational wave (GW) radiation. The merger product can only be retained in the NSC if the GW recoil kick is less than the central escape velocity of the stellar cluster which is typically of the order of a few hundred $\kms$ \citep{gerosa2019,fragione2020,gerosa2021}. Such kick velocities are systematically lower than a few hundred $\kms$ when the mass ratio of merging BHs is less than 0.15 \citepalias[see Section 6.1 in][]{askar2021}.
In a relevant recent paper, \citet{chassonnery2021} investigated whether the SMBH in the centre of the Milky Way may have formed from repeated mergers of IMBHs delivered to the galactic nuclei by stellar clusters. They carried out \textit{N}-body simulations of a dense cluster of 400 IMBHs (each having a mass of $10^{4} \ \msun$) and found that IMBHs mergers due to GW emission can efficiently occur in such a cluster leading to the formation of a seed SMBH.

In this paper, we further build on the idea that the NSC forms before the SMBH and use the results of \citetalias{askar2021} to consider the retention and growth of IMBHs in a galactic nucleus for a population of galaxies where we construct the NSC by aggregating sampled stellar clusters. This allows us to determine the observational consequences of the SMBH seeding mechanism on the demographics of NSCs and SMBHs in observed galaxies. To this end, we create a mock population of galaxies and NSCs using observed distributions and scaling relations (see Section \ref{sec:sec2-mock}). Using numerous assumptions 
based on previous works and results from \citetalias{askar2021}, each aggregated stellar cluster which forms the NSC is assigned a probability for containing an IMBH based on its mass (see Section \ref{sec:populating-galaxies}). Using this method, we estimate the number of IMBHs that can be potentially delivered by merging stellar clusters to form an NSC. We consider different cases in which we allow IMBHs to grow through GW mergers with other IMBHs and through gas accretion (see Section \ref{sec3:imbh-delivery}).

The results from this population synthesis are presented in Sections \ref{sec:results} and \ref{sec:discussion}. These results show that for high-mass NSCs, the likelihood of delivering multiple IMBHs to the NSC increases. If we account for BH growth in the NSC, the probability of retaining an SMBH seed becomes significantly higher. For low-mass galaxies (``A'' type galaxies in Fig. \ref{fig:obs-nsc-vs-bh}; e.g., M33, NGC 205) with stellar masses less than about $10^{9} \ \msun$ and NSC masses less than about $3 \times 10^{6} \ \msun$, the probability of delivering and retaining an SMBH seed in the NSC is less than 50 per cent. Therefore, the expected occupation fraction of SMBHs in these low-mass galaxies would be lower. For higher mass galaxies, with NSC masses between $10^{7}$ to $10^8 \ \msun$ (``B'' type galaxies in Fig. \ref{fig:obs-nsc-vs-bh}; e.g., MW, M31), there is a higher likelihood (between 50 to 80 per cent) of delivering and retaining an SMBH seed in the NSC. We also discuss the possible reasons why massive elliptical galaxies contain high-mass SMBHs but have lower NSC masses and a lower nucleation fraction. We also present a detailed discussion concerning the assumptions and uncertainties that go into our population synthesis calculations in Section \ref{subsec:caveats}. In Section \ref{subsec:comparison-cd2021}, we briefly compare the approach and results from \citetalias{askar2021} and this paper with the recent work by \citet{chassonnery2021}.

\section{SMBH Seed Population Synthesis}\label{sec:sec2-mock}

In order to investigate the retention and growth of SMBH seeds in NSCs based on the results presented in \citetalias{askar2021}, we carried out a population synthesis study by generating a mock sample of galaxies and their nuclei. In this section, we explain the details of this procedure and how it was used in conjunction with results from \citetalias{askar2021} to populate NSCs with IMBHs to check whether they could retain and grow an SMBH. The important steps of the procedure are described below:

\begin{itemize}
\item [1.] To create a sample of galaxy masses, we made use of the observationally-determined low-redshift galaxy mass function that is well described by a double Schechter function \citep{Baldry2012}. We sampled about $6 \times 10^{5}$ galaxies from this mass function. Stellar masses of these galaxies range between $\rm 10^{7}$ to $\rm 10^{11.5} \, \msun$.

\item [2.] For each sampled galaxy mass, we first calculate how likely that galaxy is to contain an NSC. This was done by using Fig. 3 in \citet{Neumayer2020} which illustrates the observed relation between galaxy stellar mass and its nucleation fraction. The data in Fig. 3 in \citet{Neumayer2020} was fit to several straight line segments to obtain a galaxy mass vs nucleation fraction relation. We then draw a random number between zero and one. If the drawn number is less than the nucleation fraction for that galaxy stellar mass, then we mark that galaxy as having an NSC and proceed further. Since the nucleation fraction for a given galaxy stellar mass depends on its morphological type, we assumed that 70 per cent of the galaxies we sampled were spiral and the remaining 30 per cent were elliptical \citep{Cappellari2011}.

\begin{figure}
	\includegraphics[width=\columnwidth]{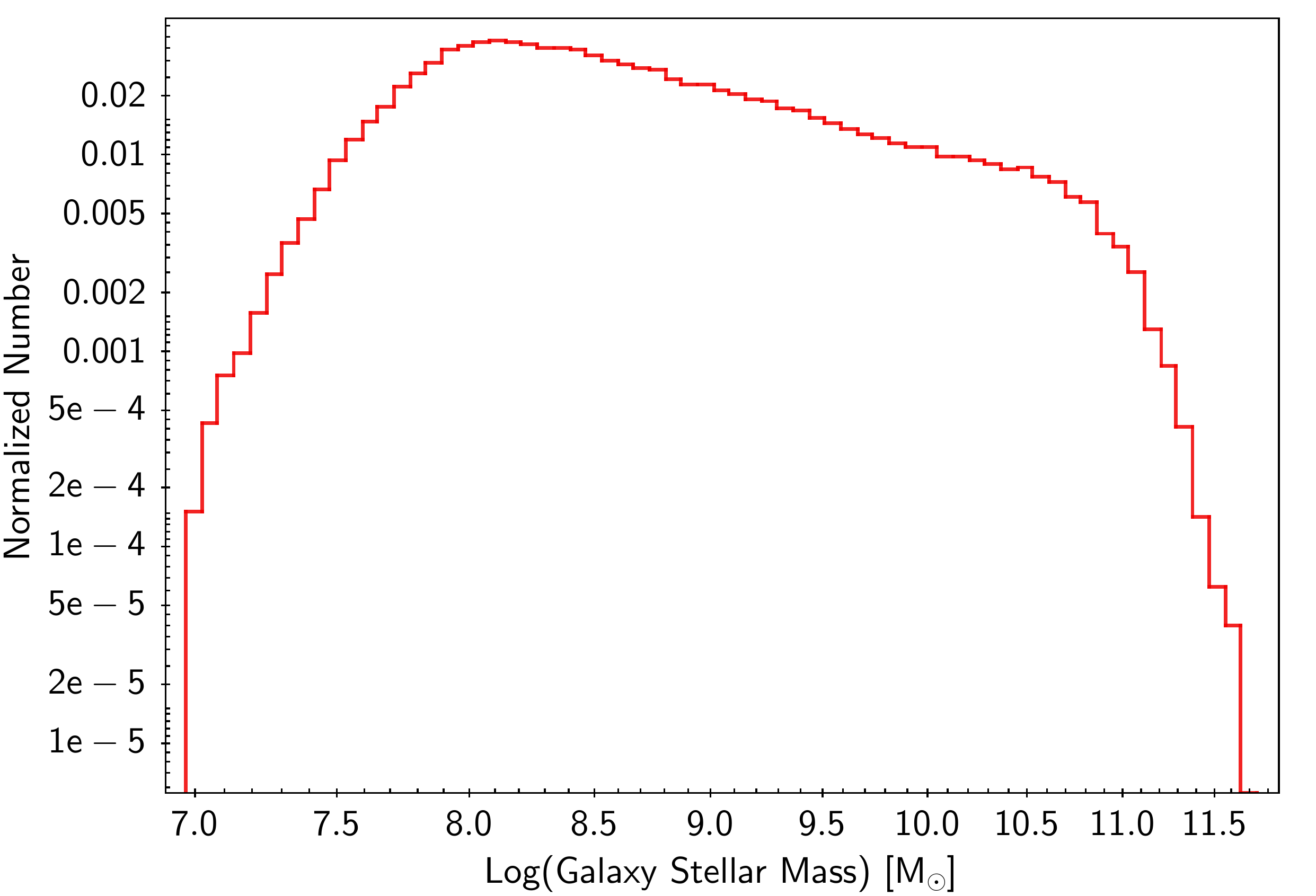}
	\caption{The distribution of galaxy stellar mass for those galaxies in our synthetic sample that contain an NSC more massive than $10^{5} \ \msun$. The number of these galaxies is $\sim 1.8 \times 10^{5}$.}
\end{figure}

\item [3.] We then calculate the corresponding NSC mass using the observed scaling relation between galaxy stellar mass ($M_{\star}$) and NSC mass ($ M_{\mathrm{NSC}}$) provided by \citet{Neumayer2020}. This relation, given in Equation \ref{eq:gal-mass-nsc-mass-relation-n2020}, was obtained by using a smaller subsample of galaxies for which accurate NSC mass measurements were available.

\begin{equation}
\log M_{\mathrm{NSC}}=0.92 \log \left(\frac{M_{\star}}{10^{9} \mathrm{M}_{\odot}}\right)+6.13
\label{eq:gal-mass-nsc-mass-relation-n2020}
\end{equation}

We also added Gaussian errors to the estimated NSC mass (Equation \ref{eq:gal-mass-nsc-mass-relation-n2020}) based on the errors provided for the fitting relations. The grey dots in Fig. \ref{fig:obs-galaxy-mass-vs-nsc-mass} show the sampled galaxy versus derived NSC masses.

\begin{figure}
    \includegraphics[width=\columnwidth]{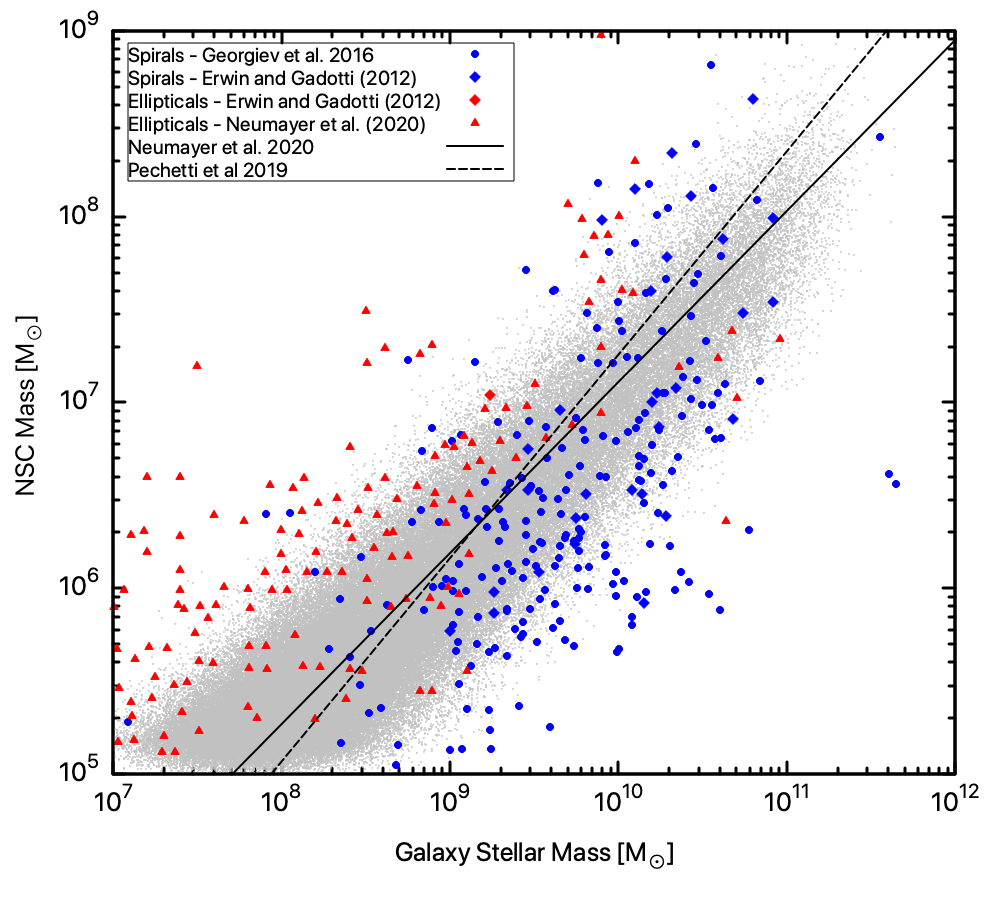}
    \caption{Galaxy stellar mass (x-axis) vs NSC mass (y-axis). The NSC masses we sampled in our population synthesis are shown as grey dots. The solid black line shows the scaling relation in Equation  \ref{eq:gal-mass-nsc-mass-relation-n2020} and the dashed black line shows the relation obtained by \citet{pechetti2020}. The blue points are observations of spiral galaxies and the red points are observations of elliptical galaxies. The blue circular points were obtained from the sample of spiral galaxies for which NSC masses were measured by \citet{Georgiev2016}. The diamond points represent galaxies for which NSC masses were obtained through dynamical and spectroscopical modelling by \citet{erwin2012}. The red triangular points are data for elliptical galaxies obtained from Fig. 12 of \citet{Neumayer2020}. These masses were derived from studies by \citet{spengler2017,ordenes2018,janssen2019}.} 

 \label{fig:obs-galaxy-mass-vs-nsc-mass}
\end{figure}

\item [4.] For each of our sampled galaxies that contains an NSC more massive than $10^{5} \ \msun$ ($\sim 1.8 \times 10^{5}$ galaxies), we draw stellar cluster masses from a log-normal distribution \citep{Schaerer2011}. The mode for this log-normal distribution is 6.41 and standard deviation is 0.52. We assume that these stellar clusters will merge in the galactic nuclei and create the NSC. So we sample clusters and aggregate their masses until their total mass is equal to the NSC mass of that galaxy. To make sure that no individually drawn stellar cluster has a mass that is significantly larger than the NSC mass, we check whether the aggregated mass exceeds the NSC mass by $5 \times 10^{4} \ \msun$  (for NSC masses less than $5 \times 10^{6} \ \msun$) and by $1 \times 10^{5} \ \msun$ (for NSC masses larger than $5 \times 10^{6} \ \msun$). Fig. \ref{fig:cluster-mass-distribution} shows the distribution of individual stellar cluster masses that were aggregated to form the NSCs larger than $10^{5} \ \msun$ . For each of the sampled stellar cluster, we also assign a random delivery time between zero and 4500 Myr. This assumes that the assembly of the NSC took place between redshifts $z \sim 4 - 1$. 

\begin{figure}
	\includegraphics[width=\columnwidth]{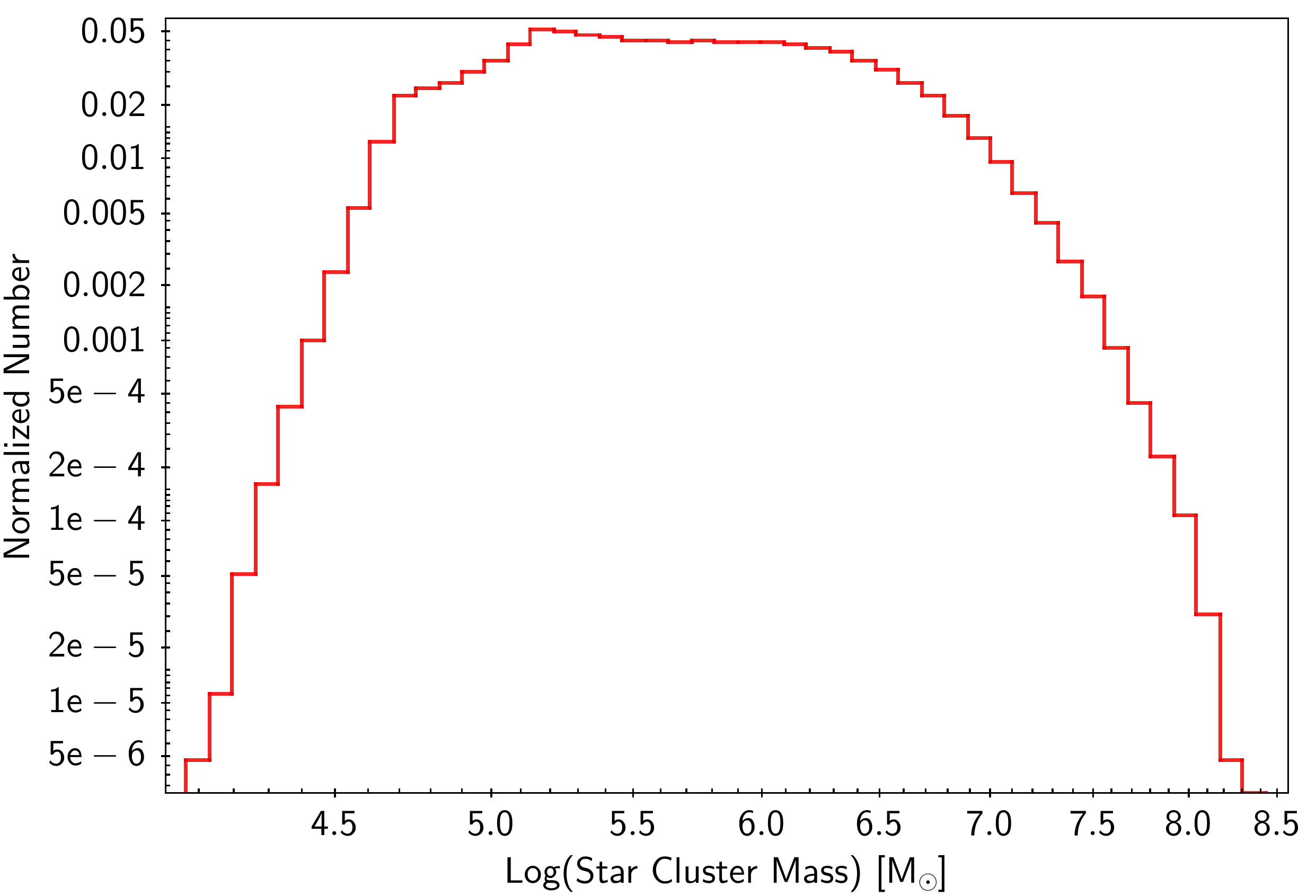}
	\caption{The distribution of the individual stellar cluster masses in our population synthesis study. These clusters were aggregated to form the NSC for galaxies that contain an NSC larger than $10^{5} \ \msun$. The masses were sampled from a log-normal distribution given by \citet{Schaerer2011}.}
	\label{fig:cluster-mass-distribution}
\end{figure}

\end{itemize}

\begin{figure*}
\resizebox{2.1\columnwidth}{!}{%

\tikzset{every picture/.style={line width=0.75pt}} 

\begin{tikzpicture}[x=0.75pt,y=0.75pt,yscale=-1,xscale=1]
\tikzstyle{every node}=[font=\large]

\draw [line width=2.25]    (251,74.83) -- (297,74.29) ;
\draw [shift={(302,74.23)}, rotate = 539.3299999999999] [fill={rgb, 255:red, 0; green, 0; blue, 0 }  ][line width=0.08]  [draw opacity=0] (14.29,-6.86) -- (0,0) -- (14.29,6.86) -- cycle    ;
\draw  [color={rgb, 255:red, 0; green, 0; blue, 0 }  ,draw opacity=1 ][fill={rgb, 255:red, 74; green, 144; blue, 226 }  ,fill opacity=0.4 ][line width=0.75]  (76.33,63.13) .. controls (76.33,58.93) and (79.74,55.53) .. (83.93,55.53) -- (243.4,55.53) .. controls (247.6,55.53) and (251,58.93) .. (251,63.13) -- (251,85.93) .. controls (251,90.13) and (247.6,93.53) .. (243.4,93.53) -- (83.93,93.53) .. controls (79.74,93.53) and (76.33,90.13) .. (76.33,85.93) -- cycle ;
\draw [line width=2.25]    (1003.5,71.5) .. controls (1093.2,71.5) and (1097.34,187.48) .. (1034.93,193.59) ;
\draw [shift={(1030,193.84)}, rotate = 359.71000000000004] [fill={rgb, 255:red, 0; green, 0; blue, 0 }  ][line width=0.08]  [draw opacity=0] (14.29,-6.86) -- (0,0) -- (14.29,6.86) -- cycle    ;
\draw  [color={rgb, 255:red, 0; green, 0; blue, 0 }  ,draw opacity=1 ][fill={rgb, 255:red, 74; green, 144; blue, 226 }  ,fill opacity=0.4 ][line width=0.75]  (302.67,55.93) .. controls (302.67,49.08) and (308.22,43.53) .. (315.07,43.53) -- (502.27,43.53) .. controls (509.11,43.53) and (514.67,49.08) .. (514.67,55.93) -- (514.67,93.13) .. controls (514.67,99.98) and (509.11,105.53) .. (502.27,105.53) -- (315.07,105.53) .. controls (308.22,105.53) and (302.67,99.98) .. (302.67,93.13) -- cycle ;
\draw [line width=2.25]    (514,74.76) -- (538,74.38) ;
\draw [shift={(543,74.3)}, rotate = 539.0799999999999] [fill={rgb, 255:red, 0; green, 0; blue, 0 }  ][line width=0.08]  [draw opacity=0] (14.29,-6.86) -- (0,0) -- (14.29,6.86) -- cycle    ;
\draw  [color={rgb, 255:red, 0; green, 0; blue, 0 }  ,draw opacity=1 ][fill={rgb, 255:red, 74; green, 144; blue, 226 }  ,fill opacity=0.4 ][line width=0.75]  (543.47,55.93) .. controls (543.47,49.08) and (549.02,43.53) .. (555.87,43.53) -- (743.07,43.53) .. controls (749.91,43.53) and (755.47,49.08) .. (755.47,55.93) -- (755.47,93.13) .. controls (755.47,99.98) and (749.91,105.53) .. (743.07,105.53) -- (555.87,105.53) .. controls (549.02,105.53) and (543.47,99.98) .. (543.47,93.13) -- cycle ;
\draw  [color={rgb, 255:red, 0; green, 0; blue, 0 }  ,draw opacity=1 ][fill={rgb, 255:red, 74; green, 144; blue, 226 }  ,fill opacity=0.4 ][line width=0.75]  (790.67,55.93) .. controls (790.67,49.08) and (796.22,43.53) .. (803.07,43.53) -- (990.27,43.53) .. controls (997.11,43.53) and (1002.67,49.08) .. (1002.67,55.93) -- (1002.67,93.13) .. controls (1002.67,99.98) and (997.11,105.53) .. (990.27,105.53) -- (803.07,105.53) .. controls (796.22,105.53) and (790.67,99.98) .. (790.67,93.13) -- cycle ;
\draw [line width=2.25]    (755.67,74.53) -- (786.33,74.36) ;
\draw [shift={(791.33,74.33)}, rotate = 539.6800000000001] [fill={rgb, 255:red, 0; green, 0; blue, 0 }  ][line width=0.08]  [draw opacity=0] (14.29,-6.86) -- (0,0) -- (14.29,6.86) -- cycle    ;
\draw  [color={rgb, 255:red, 0; green, 0; blue, 0 }  ,draw opacity=1 ][fill={rgb, 255:red, 74; green, 144; blue, 226 }  ,fill opacity=0.4 ][line width=0.75]  (772.67,174.88) .. controls (772.67,167.54) and (778.62,161.58) .. (785.97,161.58) -- (1014.37,161.58) .. controls (1021.71,161.58) and (1027.67,167.54) .. (1027.67,174.88) -- (1027.67,214.78) .. controls (1027.67,222.13) and (1021.71,228.08) .. (1014.37,228.08) -- (785.97,228.08) .. controls (778.62,228.08) and (772.67,222.13) .. (772.67,214.78) -- cycle ;
\draw [line width=2.25]    (773,194.83) -- (742.33,194.83) ;
\draw [shift={(737.33,194.83)}, rotate = 360] [fill={rgb, 255:red, 0; green, 0; blue, 0 }  ][line width=0.08]  [draw opacity=0] (14.29,-6.86) -- (0,0) -- (14.29,6.86) -- cycle    ;
\draw  [color={rgb, 255:red, 0; green, 0; blue, 0 }  ,draw opacity=1 ][fill={rgb, 255:red, 74; green, 144; blue, 226 }  ,fill opacity=0.4 ][line width=0.75]  (560.67,171.43) .. controls (560.67,162.82) and (567.65,155.83) .. (576.27,155.83) -- (722.07,155.83) .. controls (730.68,155.83) and (737.67,162.82) .. (737.67,171.43) -- (737.67,218.23) .. controls (737.67,226.85) and (730.68,233.83) .. (722.07,233.83) -- (576.27,233.83) .. controls (567.65,233.83) and (560.67,226.85) .. (560.67,218.23) -- cycle ;
\draw [line width=2.25]    (560.67,194.58) -- (533.67,195.01) ;
\draw [shift={(528.67,195.08)}, rotate = 359.1] [fill={rgb, 255:red, 0; green, 0; blue, 0 }  ][line width=0.08]  [draw opacity=0] (14.29,-6.86) -- (0,0) -- (14.29,6.86) -- cycle    ;
\draw  [color={rgb, 255:red, 0; green, 0; blue, 0 }  ,draw opacity=1 ][fill={rgb, 255:red, 74; green, 144; blue, 226 }  ,fill opacity=0.4 ][line width=0.75]  (55.67,176.23) .. controls (55.67,169.39) and (61.22,163.83) .. (68.07,163.83) -- (286.27,163.83) .. controls (293.11,163.83) and (298.67,169.39) .. (298.67,176.23) -- (298.67,213.43) .. controls (298.67,220.28) and (293.11,225.83) .. (286.27,225.83) -- (68.07,225.83) .. controls (61.22,225.83) and (55.67,220.28) .. (55.67,213.43) -- cycle ;
\draw  [color={rgb, 255:red, 0; green, 0; blue, 0 }  ,draw opacity=1 ][fill={rgb, 255:red, 74; green, 144; blue, 226 }  ,fill opacity=0.4 ][line width=0.75]  (58,323.27) .. controls (58,317.96) and (62.3,313.67) .. (67.6,313.67) -- (290.73,313.67) .. controls (296.04,313.67) and (300.33,317.96) .. (300.33,323.27) -- (300.33,352.07) .. controls (300.33,357.37) and (296.04,361.67) .. (290.73,361.67) -- (67.6,361.67) .. controls (62.3,361.67) and (58,357.37) .. (58,352.07) -- cycle ;
\draw  [color={rgb, 255:red, 0; green, 0; blue, 0 }  ,draw opacity=1 ][fill={rgb, 255:red, 74; green, 144; blue, 226 }  ,fill opacity=0.4 ][line width=0.75]  (327.67,171.43) .. controls (327.67,162.82) and (334.65,155.83) .. (343.27,155.83) -- (514.07,155.83) .. controls (522.68,155.83) and (529.67,162.82) .. (529.67,171.43) -- (529.67,218.23) .. controls (529.67,226.85) and (522.68,233.83) .. (514.07,233.83) -- (343.27,233.83) .. controls (334.65,233.83) and (327.67,226.85) .. (327.67,218.23) -- cycle ;
\draw [line width=2.25]    (327.67,194.83) -- (303.67,194.83) ;
\draw [shift={(298.67,194.83)}, rotate = 360] [fill={rgb, 255:red, 0; green, 0; blue, 0 }  ][line width=0.08]  [draw opacity=0] (14.29,-6.86) -- (0,0) -- (14.29,6.86) -- cycle    ;
\draw [line width=2.25]    (55.33,195) .. controls (-3.82,195) and (-7.51,332.52) .. (54.13,338.64) ;
\draw [shift={(59,338.84)}, rotate = 539.29] [fill={rgb, 255:red, 0; green, 0; blue, 0 }  ][line width=0.08]  [draw opacity=0] (14.29,-6.86) -- (0,0) -- (14.29,6.86) -- cycle    ;
\draw [line width=2.25]    (300.33,340.33) -- (387,339.39) ;
\draw [shift={(392,339.33)}, rotate = 539.37] [fill={rgb, 255:red, 0; green, 0; blue, 0 }  ][line width=0.08]  [draw opacity=0] (14.29,-6.86) -- (0,0) -- (14.29,6.86) -- cycle    ;
\draw [line width=2.25]    (300,338.33) .. controls (375.79,338.33) and (316.48,279.28) .. (389.37,275.66) ;
\draw [shift={(394,275.51)}, rotate = 539.05] [fill={rgb, 255:red, 0; green, 0; blue, 0 }  ][line width=0.08]  [draw opacity=0] (14.29,-6.86) -- (0,0) -- (14.29,6.86) -- cycle    ;
\draw [line width=2.25]    (313.67,339) .. controls (359.4,340.31) and (305.25,407.09) .. (383.07,412.59) ;
\draw [shift={(388,412.84)}, rotate = 182.02] [fill={rgb, 255:red, 0; green, 0; blue, 0 }  ][line width=0.08]  [draw opacity=0] (14.29,-6.86) -- (0,0) -- (14.29,6.86) -- cycle    ;
\draw  [color={rgb, 255:red, 0; green, 0; blue, 0 }  ,draw opacity=1 ][fill={rgb, 255:red, 74; green, 144; blue, 226 }  ,fill opacity=0.4 ][line width=0.75]  (393.67,259.8) .. controls (393.67,254.57) and (397.91,250.33) .. (403.13,250.33) -- (542.87,250.33) .. controls (548.09,250.33) and (552.33,254.57) .. (552.33,259.8) -- (552.33,288.2) .. controls (552.33,293.43) and (548.09,297.67) .. (542.87,297.67) -- (403.13,297.67) .. controls (397.91,297.67) and (393.67,293.43) .. (393.67,288.2) -- cycle ;
\draw  [color={rgb, 255:red, 0; green, 0; blue, 0 }  ,draw opacity=1 ][fill={rgb, 255:red, 74; green, 144; blue, 226 }  ,fill opacity=0.4 ][line width=0.75]  (392.33,327.48) .. controls (392.33,322.58) and (396.31,318.6) .. (401.21,318.6) -- (542.12,318.6) .. controls (547.02,318.6) and (551,322.58) .. (551,327.48) -- (551,354.12) .. controls (551,359.02) and (547.02,363) .. (542.12,363) -- (401.21,363) .. controls (396.31,363) and (392.33,359.02) .. (392.33,354.12) -- cycle ;
\draw  [color={rgb, 255:red, 0; green, 0; blue, 0 }  ,draw opacity=1 ][fill={rgb, 255:red, 74; green, 144; blue, 226 }  ,fill opacity=0.4 ][line width=0.75]  (388.33,395.43) .. controls (388.33,388.12) and (394.26,382.2) .. (401.56,382.2) -- (540.44,382.2) .. controls (547.74,382.2) and (553.67,388.12) .. (553.67,395.43) -- (553.67,435.11) .. controls (553.67,442.41) and (547.74,448.33) .. (540.44,448.33) -- (401.56,448.33) .. controls (394.26,448.33) and (388.33,442.41) .. (388.33,435.11) -- cycle ;
\draw [line width=2.25]    (555,416.33) .. controls (659.53,403.27) and (671.87,457.75) .. (752.01,461.19) ;
\draw [shift={(757,461.33)}, rotate = 180.9] [fill={rgb, 255:red, 0; green, 0; blue, 0 }  ][line width=0.08]  [draw opacity=0] (14.29,-6.86) -- (0,0) -- (14.29,6.86) -- cycle    ;
\draw [line width=2.25]    (553.33,416.67) .. controls (600.05,416.99) and (629.47,328.32) .. (708.44,322.27) ;
\draw [shift={(713.33,322)}, rotate = 538.15] [fill={rgb, 255:red, 0; green, 0; blue, 0 }  ][line width=0.08]  [draw opacity=0] (14.29,-6.86) -- (0,0) -- (14.29,6.86) -- cycle    ;
\draw  [color={rgb, 255:red, 0; green, 0; blue, 0 }  ,draw opacity=1 ][fill={rgb, 255:red, 74; green, 144; blue, 226 }  ,fill opacity=0.4 ][line width=0.75]  (712.67,301.8) .. controls (712.67,294.45) and (718.62,288.5) .. (725.97,288.5) -- (1021.7,288.5) .. controls (1029.05,288.5) and (1035,294.45) .. (1035,301.8) -- (1035,341.7) .. controls (1035,349.05) and (1029.05,355) .. (1021.7,355) -- (725.97,355) .. controls (718.62,355) and (712.67,349.05) .. (712.67,341.7) -- cycle ;
\draw  [color={rgb, 255:red, 0; green, 0; blue, 0 }  ,draw opacity=1 ][fill={rgb, 255:red, 74; green, 144; blue, 226 }  ,fill opacity=0.4 ][line width=0.75]  (757.67,437.8) .. controls (757.67,430.45) and (763.62,424.5) .. (770.97,424.5) -- (983.03,424.5) .. controls (990.38,424.5) and (996.33,430.45) .. (996.33,437.8) -- (996.33,477.7) .. controls (996.33,485.05) and (990.38,491) .. (983.03,491) -- (770.97,491) .. controls (763.62,491) and (757.67,485.05) .. (757.67,477.7) -- cycle ;
\draw [line width=2.25]    (876,355) -- (876.31,420.67) ;
\draw [shift={(876.33,425.67)}, rotate = 269.73] [fill={rgb, 255:red, 0; green, 0; blue, 0 }  ][line width=0.08]  [draw opacity=0] (14.29,-6.86) -- (0,0) -- (14.29,6.86) -- cycle    ;
\draw [line width=2.25]    (879,492) -- (879,559.33) ;
\draw [shift={(879,564.33)}, rotate = 270] [fill={rgb, 255:red, 0; green, 0; blue, 0 }  ][line width=0.08]  [draw opacity=0] (14.29,-6.86) -- (0,0) -- (14.29,6.86) -- cycle    ;
\draw  [color={rgb, 255:red, 0; green, 0; blue, 0 }  ,draw opacity=1 ][fill={rgb, 255:red, 74; green, 144; blue, 226 }  ,fill opacity=0.4 ][line width=0.75]  (757.67,572.87) .. controls (757.67,568.15) and (761.49,564.33) .. (766.2,564.33) -- (987.8,564.33) .. controls (992.51,564.33) and (996.33,568.15) .. (996.33,572.87) -- (996.33,598.47) .. controls (996.33,603.18) and (992.51,607) .. (987.8,607) -- (766.2,607) .. controls (761.49,607) and (757.67,603.18) .. (757.67,598.47) -- cycle ;
\draw [line width=2.25]    (999,587) .. controls (1124.4,589.64) and (1128.92,468.79) .. (1000.92,465.08) ;
\draw [shift={(997,465)}, rotate = 360.58000000000004] [fill={rgb, 255:red, 0; green, 0; blue, 0 }  ][line width=0.08]  [draw opacity=0] (14.29,-6.86) -- (0,0) -- (14.29,6.86) -- cycle    ;
\draw  [color={rgb, 255:red, 0; green, 0; blue, 0 }  ,draw opacity=1 ][fill={rgb, 255:red, 74; green, 144; blue, 226 }  ,fill opacity=0.4 ][line width=0.75]  (761.67,663.53) .. controls (761.67,658.82) and (765.49,655) .. (770.2,655) -- (991.8,655) .. controls (996.51,655) and (1000.33,658.82) .. (1000.33,663.53) -- (1000.33,689.13) .. controls (1000.33,693.85) and (996.51,697.67) .. (991.8,697.67) -- (770.2,697.67) .. controls (765.49,697.67) and (761.67,693.85) .. (761.67,689.13) -- cycle ;
\draw [line width=2.25]    (880.33,607) -- (880.33,650) ;
\draw [shift={(880.33,655)}, rotate = 270] [fill={rgb, 255:red, 0; green, 0; blue, 0 }  ][line width=0.08]  [draw opacity=0] (14.29,-6.86) -- (0,0) -- (14.29,6.86) -- cycle    ;
\draw  [color={rgb, 255:red, 0; green, 0; blue, 0 }  ,draw opacity=1 ][fill={rgb, 255:red, 74; green, 144; blue, 226 }  ,fill opacity=0.4 ][line width=0.75]  (384.33,572.87) .. controls (384.33,568.15) and (388.15,564.33) .. (392.87,564.33) -- (627.8,564.33) .. controls (632.51,564.33) and (636.33,568.15) .. (636.33,572.87) -- (636.33,598.47) .. controls (636.33,603.18) and (632.51,607) .. (627.8,607) -- (392.87,607) .. controls (388.15,607) and (384.33,603.18) .. (384.33,598.47) -- cycle ;
\draw [line width=2.25]    (879,492) -- (641.01,581.24) ;
\draw [shift={(636.33,583)}, rotate = 339.44] [fill={rgb, 255:red, 0; green, 0; blue, 0 }  ][line width=0.08]  [draw opacity=0] (14.29,-6.86) -- (0,0) -- (14.29,6.86) -- cycle    ;
\draw [line width=2.25]    (505.67,607) -- (505.67,646) ;
\draw [shift={(505.67,651)}, rotate = 270] [fill={rgb, 255:red, 0; green, 0; blue, 0 }  ][line width=0.08]  [draw opacity=0] (14.29,-6.86) -- (0,0) -- (14.29,6.86) -- cycle    ;
\draw  [color={rgb, 255:red, 0; green, 0; blue, 0 }  ,draw opacity=1 ][fill={rgb, 255:red, 74; green, 144; blue, 226 }  ,fill opacity=0.4 ][line width=0.75]  (386.33,660.2) .. controls (386.33,655.49) and (390.15,651.67) .. (394.87,651.67) -- (616.47,651.67) .. controls (621.18,651.67) and (625,655.49) .. (625,660.2) -- (625,685.8) .. controls (625,690.51) and (621.18,694.33) .. (616.47,694.33) -- (394.87,694.33) .. controls (390.15,694.33) and (386.33,690.51) .. (386.33,685.8) -- cycle ;
\draw [line width=2.25]    (636.33,583) .. controls (761.77,583.33) and (708.42,753.8) .. (779.02,759.37) ;
\draw [shift={(783.5,759.5)}, rotate = 539.01] [fill={rgb, 255:red, 0; green, 0; blue, 0 }  ][line width=0.08]  [draw opacity=0] (14.29,-6.86) -- (0,0) -- (14.29,6.86) -- cycle    ;
\draw [line width=2.25]    (1029.33,761) .. controls (1136.63,761) and (1197.72,446.83) .. (1000,446.32) ;
\draw [shift={(997,446.33)}, rotate = 359.24] [fill={rgb, 255:red, 0; green, 0; blue, 0 }  ][line width=0.08]  [draw opacity=0] (14.29,-6.86) -- (0,0) -- (14.29,6.86) -- cycle    ;
\draw [line width=2.25]    (780,760) -- (1030,760.33) ;
\draw [shift={(1035,760.33)}, rotate = 180.07] [fill={rgb, 255:red, 0; green, 0; blue, 0 }  ][line width=0.08]  [draw opacity=0] (14.29,-6.86) -- (0,0) -- (14.29,6.86) -- cycle    ;

\draw (163.67,73.33) node  [rotate=-359.85] [align=left] {\begin{minipage}[lt]{135.38392000000002pt}\setlength\topsep{0pt}
\begin{center}
Sample a galaxy stellar mass\\\citep{Baldry2012}
\end{center}

\end{minipage}};
\draw (84.67,23.33) node [anchor=north west][inner sep=0.75pt]   [align=left] {{\Large \textbf{\textcolor[rgb]{0,0,0}{Generating a mock population of galaxies with NSCs}}}};
\draw (408.67,74.33) node  [rotate=-359.85] [align=left] {\begin{minipage}[lt]{139.94196pt}\setlength\topsep{0pt}
\begin{center}
Account for nucleation fraction\\ \citep[see Fig. 3 in][]{Neumayer2020}
\end{center}

\end{minipage}};
\draw (649.47,74.53) node  [rotate=-359.85] [align=left] {\begin{minipage}[lt]{114.41pt}\setlength\topsep{0pt}
\begin{center}
Find corresponding NSC mass (see Eq. \ref{eq:gal-mass-nsc-mass-relation-n2020} and Fig. \ref{fig:obs-galaxy-mass-vs-nsc-mass}) using scaling relations \citep{Neumayer2020}
\end{center}

\end{minipage}};
\draw (896.67,74.53) node  [rotate=-359.85] [align=left] {\begin{minipage}[lt]{142.19pt}\setlength\topsep{0pt}
\begin{center}
Draw initial stellar cluster masses\\ based on \citep{Schaerer2011}
\end{center}

\end{minipage}};
\draw (893.83,192.42) node  [rotate=-359.85] [align=left] {\begin{minipage}[lt]{191.35608000000002pt}\setlength\topsep{0pt}
\begin{center}
Aggregate stellar clusters until \\ NSC Mass is reproduced (see Fig. \ref{fig:cluster-mass-distribution}) 
\end{center}

\end{minipage}};
\draw (646.67,192.33) node  [rotate=-359.85] [align=left] {\begin{minipage}[lt]{112.71000000000001pt}\setlength\topsep{0pt}
\begin{center}
Assign each cluster a\\ random delivery time\\ between \ 0 - 4500 Myr \ 
\end{center}

\end{minipage}};
\draw (428.67,194.83) node  [rotate=-359.85] [align=left] {\begin{minipage}[lt]{135.79pt}\setlength\topsep{0pt}
\begin{center}
Use probabilities in Table \ref{tab:reaslitic} to\\determine if cluster contains\\ an IMBH
\end{center}

\end{minipage}};
\draw (177.17,194.83) node  [rotate=-359.85] [align=left] {\begin{minipage}[lt]{156.92pt}\setlength\topsep{0pt}
\begin{center}
For clusters that contain an IMBH,\\ estimate IMBH mass\\ based on Table \ref{tab:sample-imbh-mass}
\end{center}

\end{minipage}};
\draw (179.17,337.67) node  [rotate=-359.85] [align=left] {\begin{minipage}[lt]{151.82088000000002pt}\setlength\topsep{0pt}
\begin{center}
Sort delivered clusters and IMBH\\ by time 
\end{center}

\end{minipage}};
\draw (650.67,256.71) node [anchor=north west][inner sep=0.75pt]  [rotate=-0.35] [align=left] {{\large \textbf{Retention and growth of seed SMBHs}}};
\draw (473,274) node  [rotate=-359.85] [align=left] {\begin{minipage}[lt]{106.47304000000001pt}\setlength\topsep{0pt}
\begin{center}
No IMBH delivered:\\no SMBH seed in NSC
\end{center}

\end{minipage}};
\draw (471.67,340.8) node  [rotate=-359.85] [align=left] {\begin{minipage}[lt]{92.28892pt}\setlength\topsep{0pt}
\begin{center}
1 IMBH delivered:\\SMBH seed in NSC
\end{center}

\end{minipage}};
\draw (471,415.27) node  [rotate=-359.85] [align=left] {\begin{minipage}[lt]{111.57304pt}\setlength\topsep{0pt}
\begin{center}
For multiple IMBH: \\Prompt mergers between\\ IMBH pairs 
\end{center}

\end{minipage}};
\draw (577.21,364.59) node [anchor=north west][inner sep=0.75pt]  [rotate=-326.6,xslant=-0.06] [align=left] {\textbf{Growth of BHs }};
\draw (873.83,321.75) node  [rotate=-359.85] [align=left] {\begin{minipage}[lt]{204.56304pt}\setlength\topsep{0pt}
\begin{center}
Allow IMBH to grow during the time it takes\\for the next IMBH to be delivered to the NSC
\end{center}

\end{minipage}};
\draw (876.5,457.75) node  [rotate=-359.85] [align=left] {\begin{minipage}[lt]{138.21pt}\setlength\topsep{0pt}
\begin{center}
Check mass ratio (q) between\\merging IMBH pairs
\end{center}

\end{minipage}};
\draw (891.33,509.33) node [anchor=north west][inner sep=0.75pt]   [align=left] {\textbf{If mass ratio (q) is < \ 0.15}};
\draw (877,585.67) node  [rotate=-359.85] [align=left] {\begin{minipage}[lt]{142.18392pt}\setlength\topsep{0pt}
\begin{center}
Merged BH is retained in NSC:\\ sum up IMBH masses
\end{center}

\end{minipage}};
\draw (881,676.33) node  [rotate=-359.85] [align=left] {\begin{minipage}[lt]{146.12588pt}\setlength\topsep{0pt}
\begin{center}
\textbf{1 SMBH seed retained in NSC}
\end{center}

\end{minipage}};
\draw (584.13,424.05) node [anchor=north west][inner sep=0.75pt]  [rotate=-19.91] [align=left] {\textbf{No Growth of BHs }};
\draw (507,585) node  [rotate=-359.85] [align=left] {\begin{minipage}[lt]{164.85784pt}\setlength\topsep{0pt}
\begin{center}
Merged BH is ejected from the NSC
\end{center}

\end{minipage}};
\draw (641,553.99) node [anchor=north west][inner sep=0.75pt]  [rotate=-341.08] [align=left] {\textbf{If mass ratio (q) is > \ 0.15}};
\draw (505.67,673) node  [rotate=-359.85] [align=left] {\begin{minipage}[lt]{154.05196pt}\setlength\topsep{0pt}
\begin{center}
\textbf{No SMBH seed retained in NSC}
\end{center}

\end{minipage}};

\end{tikzpicture}

}
\caption{This flowchart is a schematic diagram of the main steps involved in creating our mock population of galaxies with NSCs and seed IMBHs. Starting from the top left side where we first sample a galaxy stellar mass and then find the corresponding NSC mass after accounting for the nucleation fraction (see points one, two and three in Section \ref{sec:sec2-mock} for details). Stellar cluster masses are sampled and aggregated until we can reproduce the NSC mass. Each stellar cluster is assigned a random delivery time between 0 and 4500 Myr. Depending on its mass, the stellar cluster may or may not deliver an IMBH to the NSC (see Section \ref{sec:populating-galaxies}). In the lower right half of the figure, the different outcomes for SMBH seed formation in the NSC after accounting for IMBH retention and growth (see Section \ref{sec3:imbh-delivery}) are illustrated .}
   \label{fig:flow-chart}
\end{figure*}
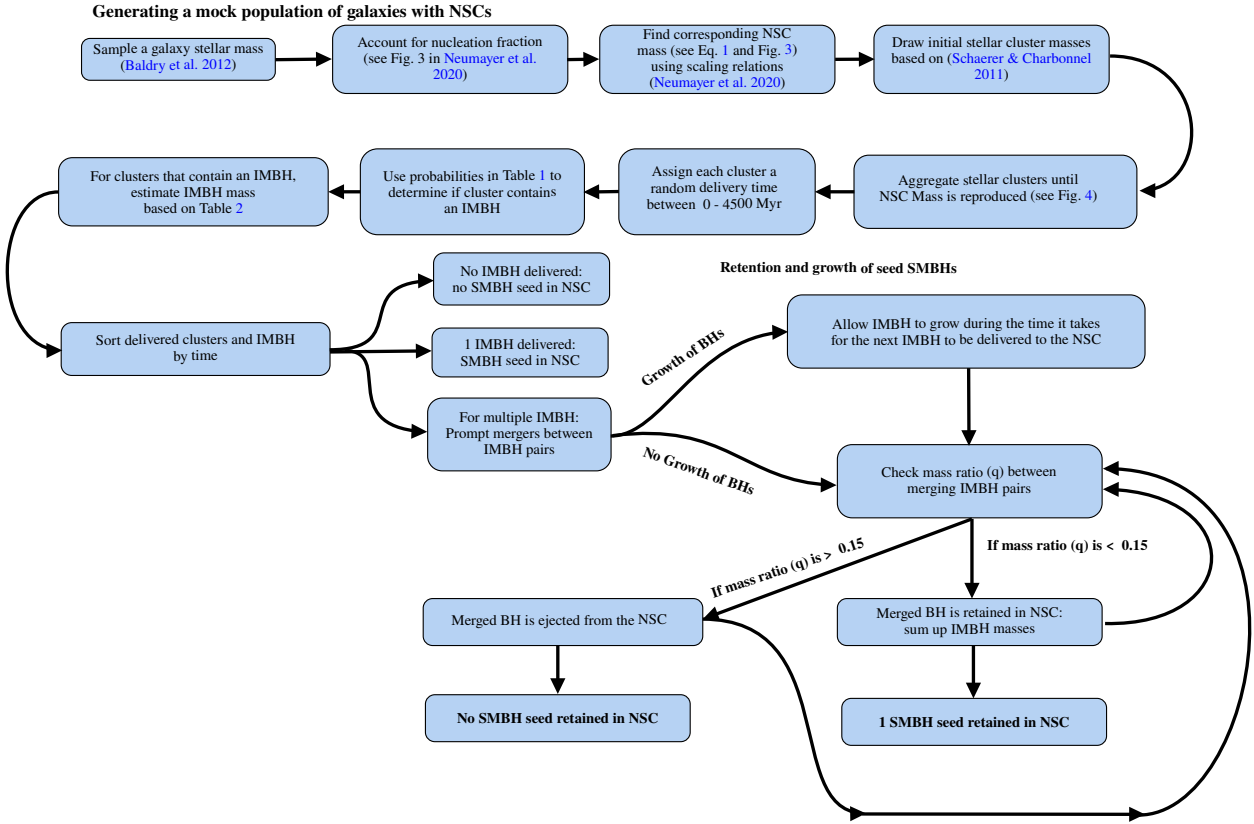

\subsection{Populating galactic nuclei with IMBHs}\label{sec:populating-galaxies}



As discussed in detail in Section 3.1 of \citetalias{askar2021}, several studies have hypothesized that IMBHs of $10^{2} - 10^{4}\,\rm \msun$ can form in dense stellar clusters \citep[see][and references therein]{greene2019rev}. During the early evolution of dense stellar clusters, runaway mergers between massive stars may result in the formation of a very massive star \citep{freitag2006,mapelli2016,gieles2018,reinoso2018}. It may be possible for this massive star to evolve into an IMBH seed \citep{gurkan2004,seguel2020}, particularly in low metallicity environments where mass loss due to stellar winds is low. The BH seed can then grow through subsequent mergers with surrounding stars   \citep{tagawa2020,das2021b}. An IMBH may also form through the gradual growth of stellar-mass BHs and their progenitors via mergers with other BHs or stars \citep{rizutto2020,gonzalez2021,dicarlo2021}. Moreover, stellar-mass seed BHs can also grow by accretion of gas in primordial, gas-rich, massive stellar clusters \citep{vesperini2010,leigh2013}. While there is no clear evidence for IMBHs in Galactic globular clusters (see Section \ref{subsec-caveats:imbh-formation-growth} for details), the interpolation of observed host mass versus BH mass correlation for galaxies to globular cluster masses would be consistent with their presence (see Fig.\,1 in \citet{chassonnery2021}).

\begin{figure}
	 \includegraphics[width=\columnwidth]{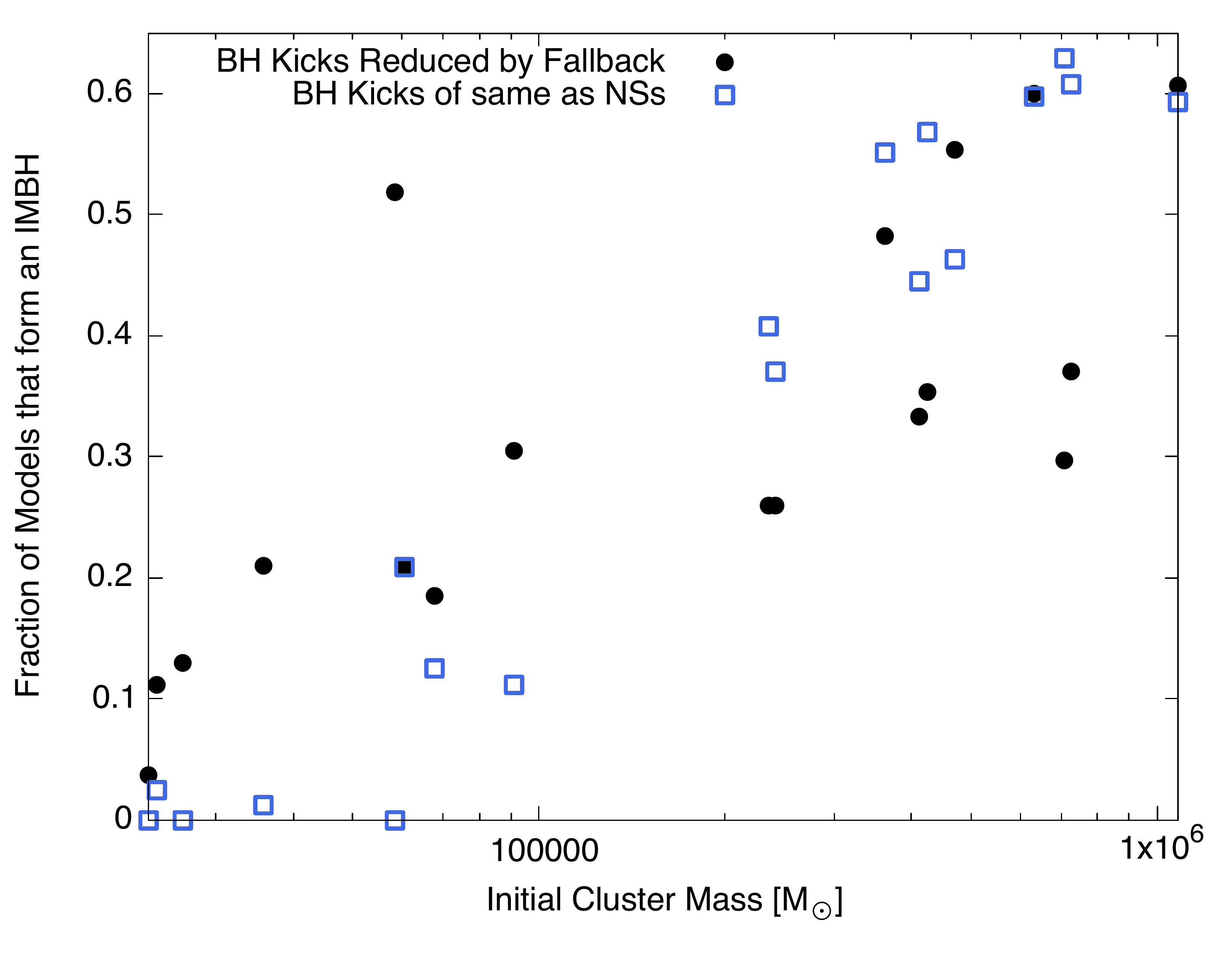}
	\caption{The fraction of models that form an IMBH in \textsc{mocca}-Survey Database I as a function of the initial stellar cluster mass. The black circles are models in which BH natal kicks were scaled by the fallback prescription from \citet{belczynski2002}. Blue squares represent models where BH natal kicks were the same as those for neutron stars. These kicks follow a Maxwellian distribution with $\sigma = 265 \ \kms$  based on observations by \citet{hobbs2005}.}
    \label{fig:gc-mass-imbh}
\end{figure}

The \textsc{mocca}-Survey Database I \citep{askar2017} comprises around 2000 stellar cluster models with different initial parameters simulated using the \textsc{mocca} code \citep{hypki2013,giersz2013}. Among these models, about 400 form an IMBH which is more massive than $100 \ \msun$ through the processes described in \citet{giersz2013,Sedda2019,hong2020}. Using the results from these simulations, we determined what fraction of models for a given initial stellar cluster mass can form an IMBH. Fig. \ref{fig:gc-mass-imbh} shows the fraction of models that form an IMBH larger than $100 \ \msun$ as a function of the initial cluster mass. We divide the \textsc{mocca}-Survey Database I models in two categories: one for models in which the natal kicks for stellar-mass BHs are the same as for NS and follow a Maxwellian distribution with $\sigma = 265\,\text{km}\text{s}^{-1}$ \citep{hobbs2005} (high natal kicks); and the other where the masses and natal kicks for stellar BHs are computed using the mass fallback prescriptions of \citet{belczynski2002} (low natal kicks).

We find that for clusters more massive than a few $10^{5} \ \msun$, the fraction of models that form an IMBH can be between 25-50 per cent depending on their initial parameters. However, the models simulated in \textsc{mocca}-Survey Database I \citep{askar2017} do not account for GW recoil kicks following the merger of two BHs. Such recoil kicks can inhibit the formation of an IMBH by ejecting the merged BH from the cluster. It was shown in \citet{morawski2018} that accounting for GW recoil kicks would result in the formation of an IMBH in only about 25 per cent of the \textsc{mocca}-Survey Database I simulations. Additionally, in the \citet{askar2017} simulations, the merger of a star and a BH results in the complete accretion of the star by the BH. \citet{Giersz15} showed that reducing the amount of accreted mass to 25 per cent does not significantly affect the growth of the IMBH. Given that all these assumptions in the simulated models are conducive to forming IMBHs, we treat the results shown in Fig. \ref{fig:gc-mass-imbh} as upper limits for IMBH formation at a given cluster mass and scale down the results of Fig.~\ref{fig:gc-mass-imbh} when determining whether a cluster can form an IMBH. In particular, we impose that for the most massive clusters ($10^{7} \ \msun$), one in every three clusters could form an IMBH. We show the probabilities that we adopt for the formation of an IMBH at a given initial stellar cluster mass in Table \ref{tab:reaslitic}. The veracity of the assumptions and observational constraints concerning IMBH formation in stellar clusters is further discussed in Section \ref{subsec-caveats:imbh-formation-growth}.



\begin{table}
\centering
\caption{Probability whether a sampled star mass contains an IMBH. These probabilities are based on the results of nearly two thousand cluster models simulated using \textsc{mocca}-Survey Database I \citep{askar2017}. We consider two cases, one is based on results from models in which BHs had natal kicks as high as NSs (High natal kicks) and the other are based on results from models in which BHs had natal kicks lowered due to fallback (Low natal kicks).}
\label{tab:reaslitic}
\begin{tabular}{cccc}
\hline
Initial Mass of Stellar & \multicolumn{3}{c}{Probability of IMBH Formation} \\ Cluster [$\msun$] & \\ \hline
\multicolumn{1}{l}{} & \multicolumn{1}{l}{High natal kicks} & \multicolumn{1}{l}{Low natal kicks} \\
M > $\rm 10^{7}$ & 0.33 & 0.33\\
$10^{6}$ < M < $10^{7}$ & 0.33 & 0.25\\
$5 \times10^{5}$ < M < $10^{6}$ & 0.25 & 0.25\\
$1 \times10^{5}$ < M < $5 \times10^{5}$ & 0.13 & 0.16 \\
M < $\rm 10^{5}$ & 0.0 & 0.14
\end{tabular}
\end{table}

\begin{figure}
	 \includegraphics[width=\columnwidth]{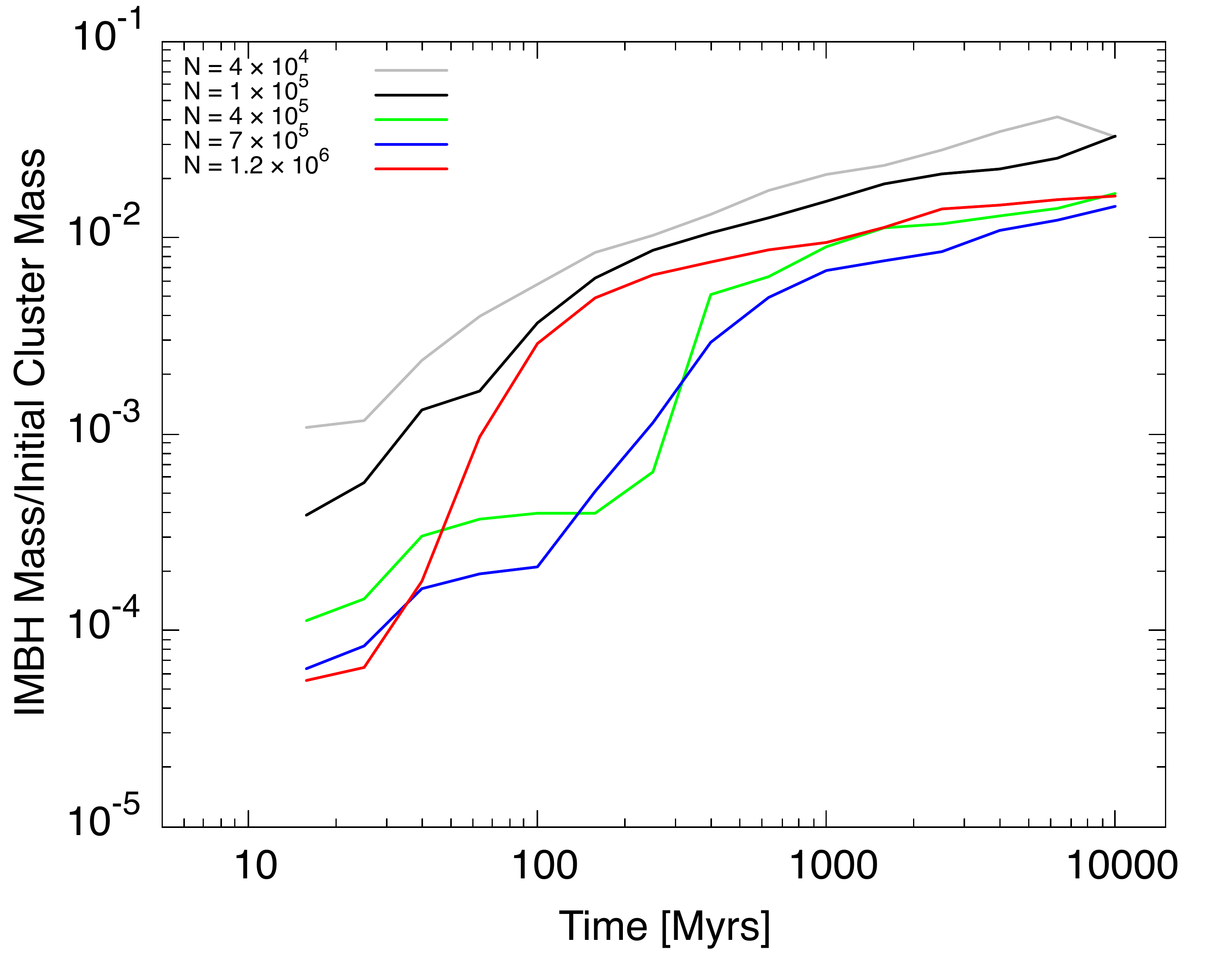}
	\caption{Evolution of the ratio between IMBH mass and initial cluster mass for \textsc{mocca}-Survey Database I with different initial number of objects that form an IMBH. The grey line is for the lowest mass stellar clusters with $4 \times 10^{4}$ initial number of objects. The black, green, blue, and red lines respectively correspond to stellar clusters with $1 \times 10^{5}$, $4 \times 10^{5}$, $7 \times 10^{5}$ and $1.2 \times 10^{6}$ initial number of objects.}
    \label{fig:time-vs-frac}
\end{figure}

Using the probabilities given in Table \ref{tab:reaslitic}, we randomly determine whether a sampled stellar cluster contains an IMBH or not (see point 4 in Section \ref{sec:sec2-mock}). We also assign masses to the IMBHs that could potentially form in the clusters. For this purpose, we again make use of results from the \textsc{mocca}-Survey Database I in order to check the evolution of the ratio between the IMBH mass and initial cluster mass for models with different initial numbers of stars. This evolution is shown in Fig. \ref{fig:time-vs-frac}. Using these results, we assume that the clusters can infall anytime between a few tens to up to a few thousand Myr. Thus for a given initial GC Mass, we estimate the mass of its IMBH by randomly sampling between the ranges provided in Table \ref{tab:sample-imbh-mass}. The IMBH mass delivered to the NSC is at most one per cent of the initial cluster mass. Through this probabilistic sampling, we can estimate how many IMBHs could potentially be delivered to an NSC.

\begin{table}
\centering
\caption{Sampled IMBH mass for a given initial stellar cluster mass.}
\label{tab:sample-imbh-mass}
\begin{tabular}{cc}
\hline
Initial Mass of Stellar & Sampled IMBH Mass as \\  Cluster [$\msun$] & Fraction of Initial Cluster Mass  \\ \hline
M > $\rm 10^{7}$ & 0.0001 to 0.009 \\
$10^{6}$ < M < $10^{7}$ & 0.0001 to 0.01 \\
$5 \times10^{5}$ < M < $10^{6}$ & 0.0002 to 0.01 \\
$1 \times10^{5}$ < M < $5 \times10^{5}$ & 0.0003 to 0.01 \\
M < $\rm 10^{5}$ & 0.0003 to 0.015
\end{tabular}
\end{table}

\section{IMBH delivery and retention}\label{sec3:imbh-delivery}

\subsection{IMBH delivery}

For NSCs with higher mass ($\rm M \gtrsim 10^{7} \ \msun$), more stellar clusters need to be aggregated in order to account for their mass. This increases the likelihood that one of those clusters can bring along an IMBH (see Table \ref{tab:reaslitic}) to the galactic nuclei.
From the sampling procedure described in Section \ref{sec:sec2-mock}, we calculated the distribution of the number of IMBHs accreted by our sampled NSCs for different ranges of NSC mass. The stacked histogram in Fig. \ref{fig:histo-delivered-low} shows the fraction of NSCs that accreted zero, one, two or more IMBHs. For the sampling that was done with both the high and low natal kick assumptions (see Table \ref{tab:reaslitic}), we find that NSCs with masses less than $10^{6} \ \msun$ are more likely to have no IMBH delivered. Typically, such NSCs do not require many stellar clusters to build their mass and therefore the likelihood of delivering more than one IMBH is extremely low. For NSCs in the mass range $10^{6}$ to $10^{7} \ \msun$, the likelihood of delivering no IMBH reduces to less than 50 per cent. Our sampling shows that a high fraction of these NSCs can have one or two IMBHs delivered to them.

For each delivered IMBH, we have randomly assigned the time at which it was delivered to the galactic nucleus. As explained earlier, this delivered time is sampled between zero and 4500 Myr. This gives us the chronological order in which the IMBH are delivered to the assembling NSC. In the case where no IMBH is delivered, we assume that there will be no SMBH seed in the NSC and thus the galaxy will not harbour an SMBH. If a single IMBH is delivered to the NSC then there will be an SMBH seed. In Section \ref{subsec:imbh-retention}, we discuss the treatment for the retention and growth of an SMBH seed when multiple IMBHs are delivered to the NSC.

\subsection{IMBH Retention}\label{subsec:imbh-retention}

If more than one IMBH is delivered to an NSC in our sampling procedure, based on results from \citetalias{askar2021}, it is assumed that the IMBHs will sink to the centre of the NSC and form a binary system . For simplicity, we assume that the two IMBHs in the binary system will promptly merge following the delivery of the second IMBH. The retention of the merged BH will depend on the magnitude of the GW recoil kick which it receives. These kicks depend on the mass ratio of the merging BH, and the magnitude and orientation of their spins. Taking the BH spin magnitude distributions from \citet{Lousto2012}, it was argued in \citetalias[][(see Fig. 16 and Section 6.1)]{askar2021} that the merged BH can be retained in typical NSCs provided that the mass ratio of the merging IMBH is low ($\rm q \lesssim 0.15$), otherwise GW recoil kicks are more likely to be larger than a few hundred $\kms$. If the mass ratio of the merging BHs is greater than 0.15, the merged BH is unlikely to be retained within the NSC. Applying these results to this work, we only retain the merged BH in cases where the mass ratio between the first and second merging BH is less than 0.15. This procedure is applied to each successive pair of merging BHs in our sampled NSC until one or no IMBH remains within the NSC. The 0.15 mass ratio threshold taken here is based on the median values for GW recoil kicks from the calculations done in \citetalias{askar2021}. We note that this calculation is sensitive to the assumed spin magnitude distribution of BHs. With lower spin values, it may be possible to retain merged BHs that have a higher mass ratios \citep{chassonnery2021,2021arXiv211109223M}. The influence of this assumption on the outcome of the results is further discussed in Section \ref{subsec:gw-recoil-kick}.


We also consider the case where the IMBHs may grow through accretion of gas from the time they were delivered to the NSC up to the delivery of the next IMBH. Accounting for this growth has an important consequence on the retention of the merged BH following the GW merger of a binary IMBH.
The e-folding time for a BH accreting at the Eddington rate is roughly 30-50 Myr \citep{madau2014}. If we assume that the BH accretes at 10 per cent of the Eddington rate during the time it can grow, then the mass doubling time for it would be about 300 Myr. This average growth rate for BHs is roughly consistent with growth histories of SMBHs derived from luminosity functions of active galaxies and results in present-day SMBH masses that match observations. A more detailed discussion on the motivation and validity of this assumption is provided in Section \ref{ad-hoc-bh-growth}. Using this doubling time and the time difference between the IMBH delivery to the NSC and the delivery of the next IMBH, we can estimate the growth in the IMBH mass using the following equation
\begin{equation}
M=M_{0} \times 2^{t / \tau}
\label{eq:growth}
\end{equation}
where $M_{0}$ is the initial BH mass, $\rm \tau$ is the doubling time which is set to 300 Myr, and $t$ is the time difference between the delivery of the first IMBH and the next IMBH to the NSC. We then assume that the pair of IMBHs will merge and the merged BH can be retained if the mass ratio is less than 0.15. Otherwise, the merged BH is ejected. This procedure is repeated until one or no BH is left in the NSC. If a BH remains in the cluster at the end of this procedure, it is grown from the time of the last BH merger or IMBH delivery to 4500 Myr. Similarly, where only one IMBH is delivered to the NSC. we grow the IMBH from the time it was delivered to 4500 Myr. Therefore, to find $\rm t$ in Equation \ref{eq:growth} in the case of one IMBH, we subtract the delivery time of the IMBH from 4500 Myr. As discussed in Section \ref{sec:sec2-mock}, our procedure assumes that the NSC assembly takes place between redshifts $1<z<4$ and the age difference between these redshifts is about 4500 Myr. Hence it is taken as an upper limit for the time up to which the BH can grow.




\section{Results}\label{sec:results}

\begin{figure}
     \includegraphics[width=\columnwidth,scale=1]{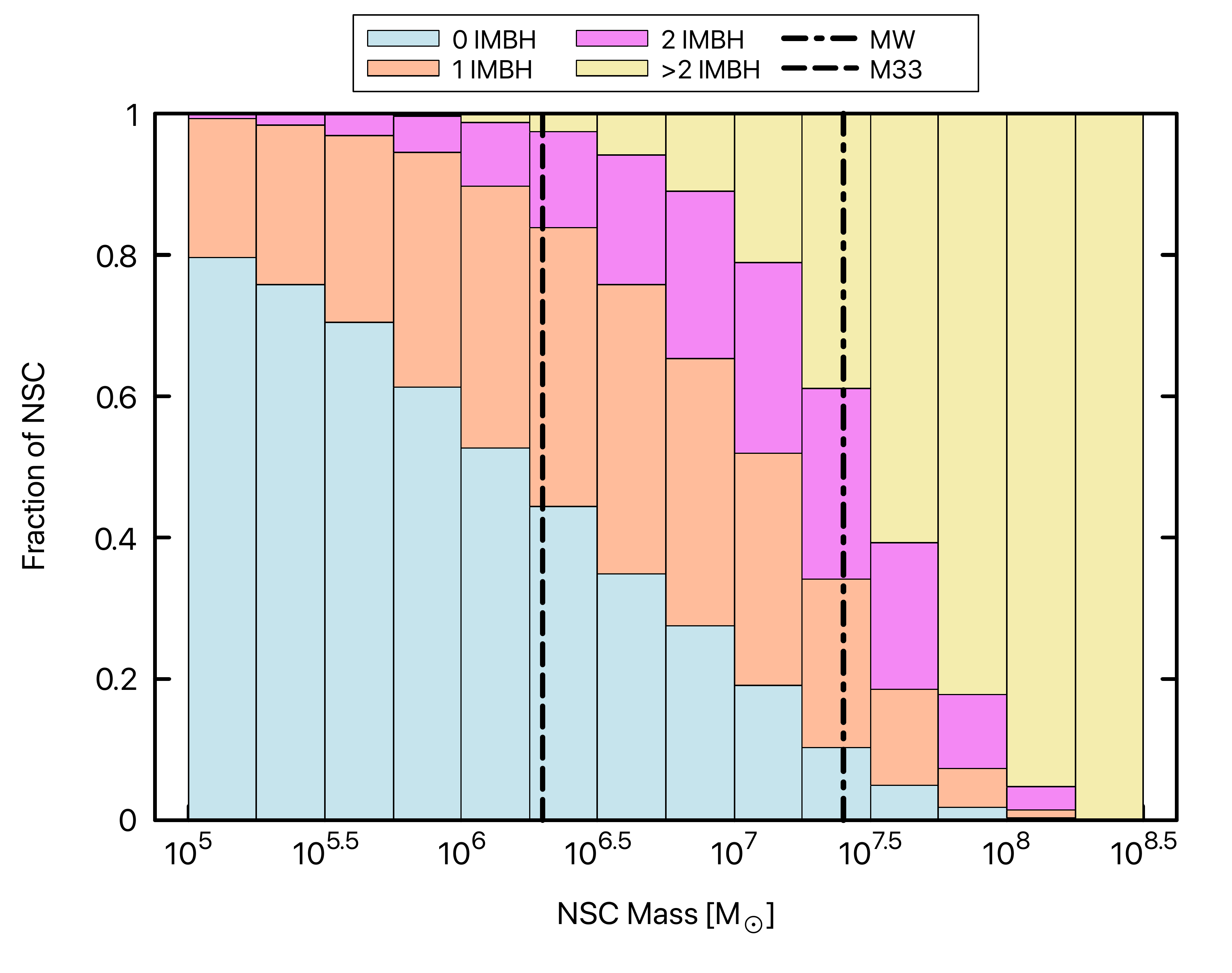}
	\caption{The fraction of NSCs with different numbers of delivered IMBHs ($y$-axis) corresponding to different NSC mass bins ($x$-axis). We assume a low natal kick for stellar-mass BHs when determining probability of IMBH formation (see Table \ref{tab:reaslitic}). It can be seen that for NSC masses less than $\sim 10^{6} \ \msun$, the likelihood of delivering no IMBHs (shown in light blue) is high. The likelihood of delivering one (orange), two (pink) or more IMBHs (yellow) increases with NSC mass. The dashed vertical line indicates the mass of the NSC in M33 while the dashed-dotted line shows the mass of the NSC in the Milky Way.}
    \label{fig:histo-delivered-low}
\end{figure}

\begin{figure*}
 	  \includegraphics[width=0.49\linewidth,scale=1]{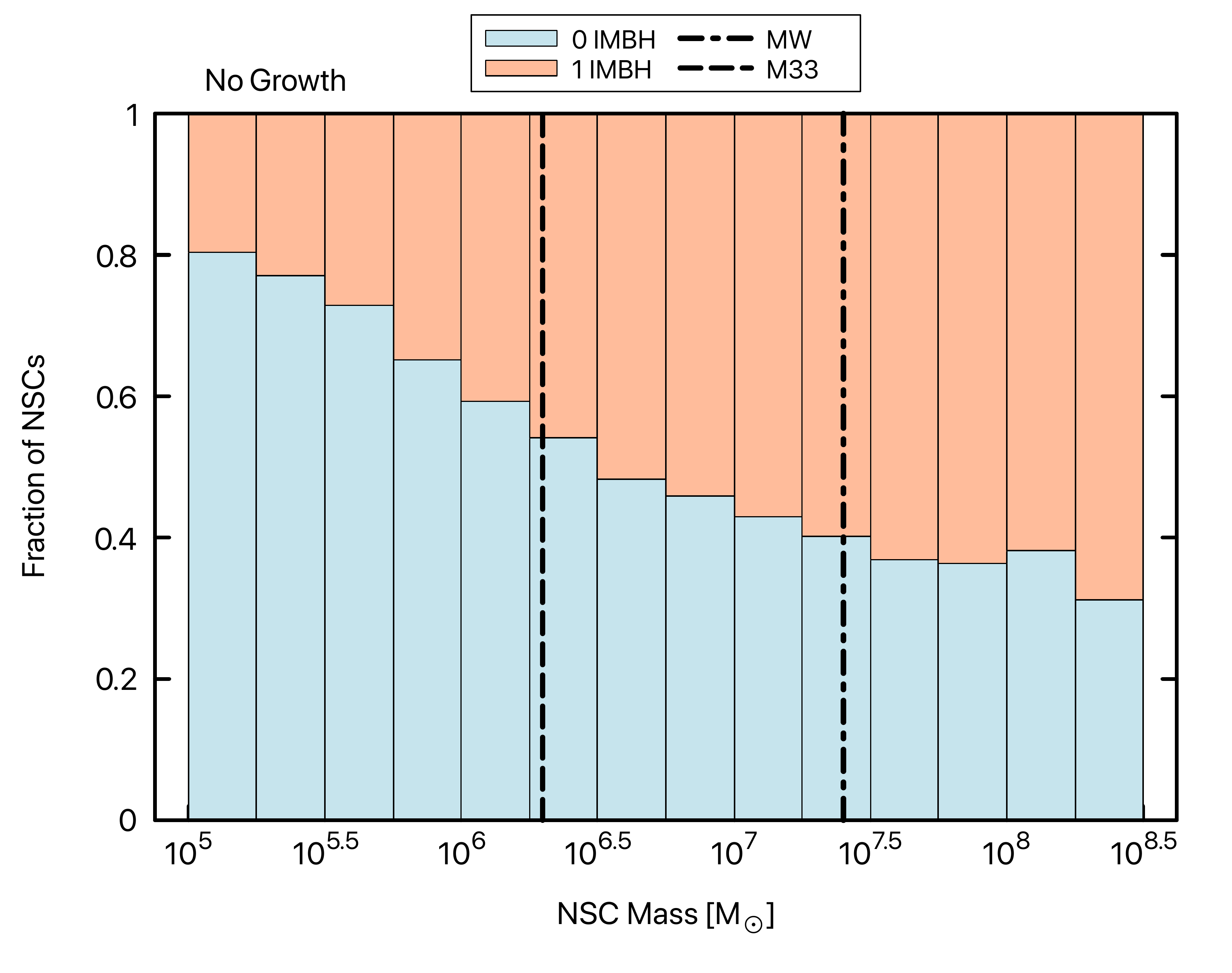}
 	  \includegraphics[width=0.49\linewidth,scale=1]{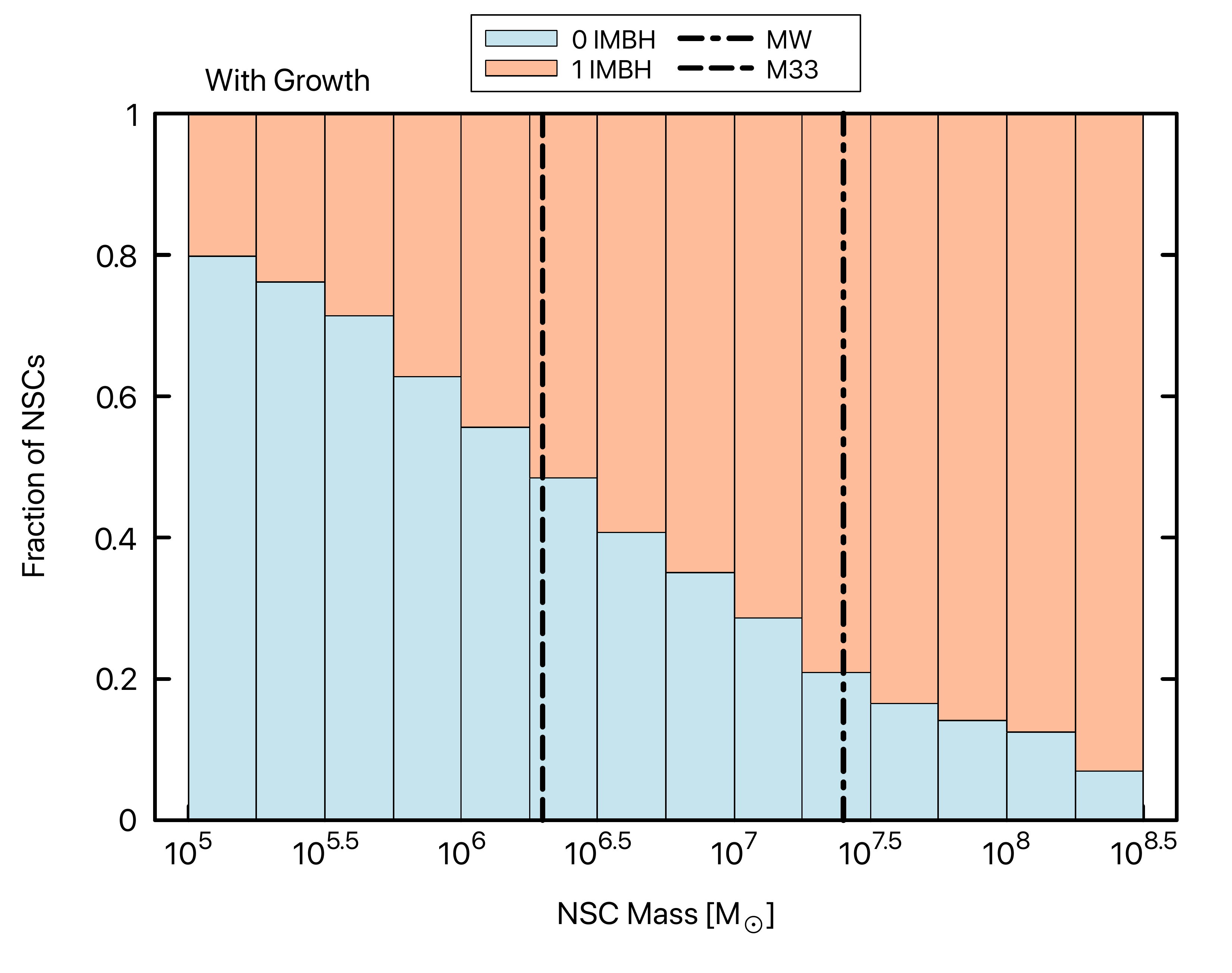}
	\caption{Retention of an SMBH seed in the NSC for different assumptions about BH growth. The left panel shows the case where we do not consider BH growth and only check the mass ratio of the merging BHs when more than one IMBH is delivered to the NSC. The right panel shows the case where we allow for growth of the central BH with a mass doubling time of 300 Myr. The fraction of NSCs in which either no IMBH was delivered, or all delivered IMBHs were subsequently ejected, is shown in light blue; the fraction of NSCs in which the SMBH seed is retained is shown in orange.}
    \label{fig:histo-processed}
\end{figure*}

In this section, we present the results from our population synthesis approach described in Sections \ref{sec:sec2-mock} and \ref{sec3:imbh-delivery}. For simplicity, we focus on results from the case where we assumed lower natal kicks for BHs when determining the probability of IMBH formation for a given stellar cluster in Table \ref{tab:reaslitic}. The stacked histogram in Fig. \ref{fig:histo-delivered-low} shows the fraction of NSCs in which zero, one, two or more IMBHs were delivered as a function of NSC mass. It can be clearly seen that NSCs with a mass less than $10^{6} \ \msun$ are highly unlikely to have any IMBH delivered to their NSCs. For these low NSC masses, only a few relatively low-mass stellar clusters need to be aggregated in order to form the NSC and the probability of any one of those clusters to have formed an IMBH  which can potentially be a seed for the SMBH is low.

For NSC masses between $10^{6}$ to $10^{7.5} \ \msun$, the probability of delivering at least one IMBH goes up significantly with NSC mass and the likelihood of no IMBH being delivered to the NSC decreases significantly to less than about 10 per cent for NSCs more massive than $10^{7} \ \msun$. The results from Fig. \ref{fig:histo-delivered-low} show that multiple IMBHs can be delivered to the NSC in galaxies where the NSC mass is larger than $\rm 10^{7} \ \msun$. Given that many of these NSCs are likely to be assembled by at least a few massive stellar clusters, the IMBH being delivered by the more massive stellar cluster is likely to be more massive (see Fig. \ref{fig:gc-mass-imbh} and Table \ref{tab:sample-imbh-mass}). On the other hand, lower mass stellar clusters are more likely to deliver less massive IMBHs. In Fig. \ref{fig:heatmap-all} in Appendix \ref{appendixA}, we have also presented the main results shown in Fig. \ref{fig:histo-delivered-low} as heatmaps. In the left panel of Fig. \ref{fig:heatmap-all}, we also show the results for the case where we assumed a higher natal kick for stellar-mass BHs when determining the probability of IMBH formation.

\subsection{Consequences of IMBH mergers on their growth and retention in the NSC}\label{subsec:processed}

If a single IMBH is delivered to the NSC then it may grow and become an SMBH. However, as outlined in Section \ref{subsec:imbh-retention}  when multiple IMBHs are delivered, the retention of the IMBH following the merger will depend on the mass ratio of the merging BHs. With the procedure described in Section \ref{subsec:imbh-retention}, a prompt merger for an IMBH binary is assumed and we are either left with none or one IMBH in the end. The stacked histograms in Fig. \ref{fig:histo-processed} shows the fractions of NSCs in which either zero or one IMBH remains at the end for two different cases:
\begin{enumerate}
    \item \textit{No Growth} (left panel in Fig. \ref{fig:histo-processed}): In this case, we do not allow for BH growth between the time of its delivery to the NSC and the delivery of the next IMBH. Following the merger of an IMBH binary, the merged BH is only retained if the mass ratio of the merging BHs is less than 0.15, otherwise the merged BH is removed from the system. This assumes that for mass ratios higher than 0.15, the recoil kick would be larger than the central escape speed of the NSC.
    \item \textit{BH Growth with mass doubling time of 300 Myr} (right panel in Fig. \ref{fig:histo-processed}): In this case, we allow the first delivered BH to grow with a doubling time of 300 Myr from the time of its delivery up to the time the next IMBH is delivered. This assumes that the BH mass accretion rate will be about 10 per cent of the Eddington rate over the time the BH spends in the galactic nuclei from the time of its delivery up to redshift 1.
\end{enumerate}

Additionally we also considered a third case, \textit{BH Growth at Eddington rate}, where we assume that the BH can accrete at its Eddington rate with a mass doubling time of 30 Myr. While it is unlikely that the BH will be able to accrete and grow at this rate over several hundred to a few thousand Myr, we used this extreme growth rate to investigate the influence of IMBH retention following GW mergers.

\begin{figure}
	 \includegraphics[width=\columnwidth]{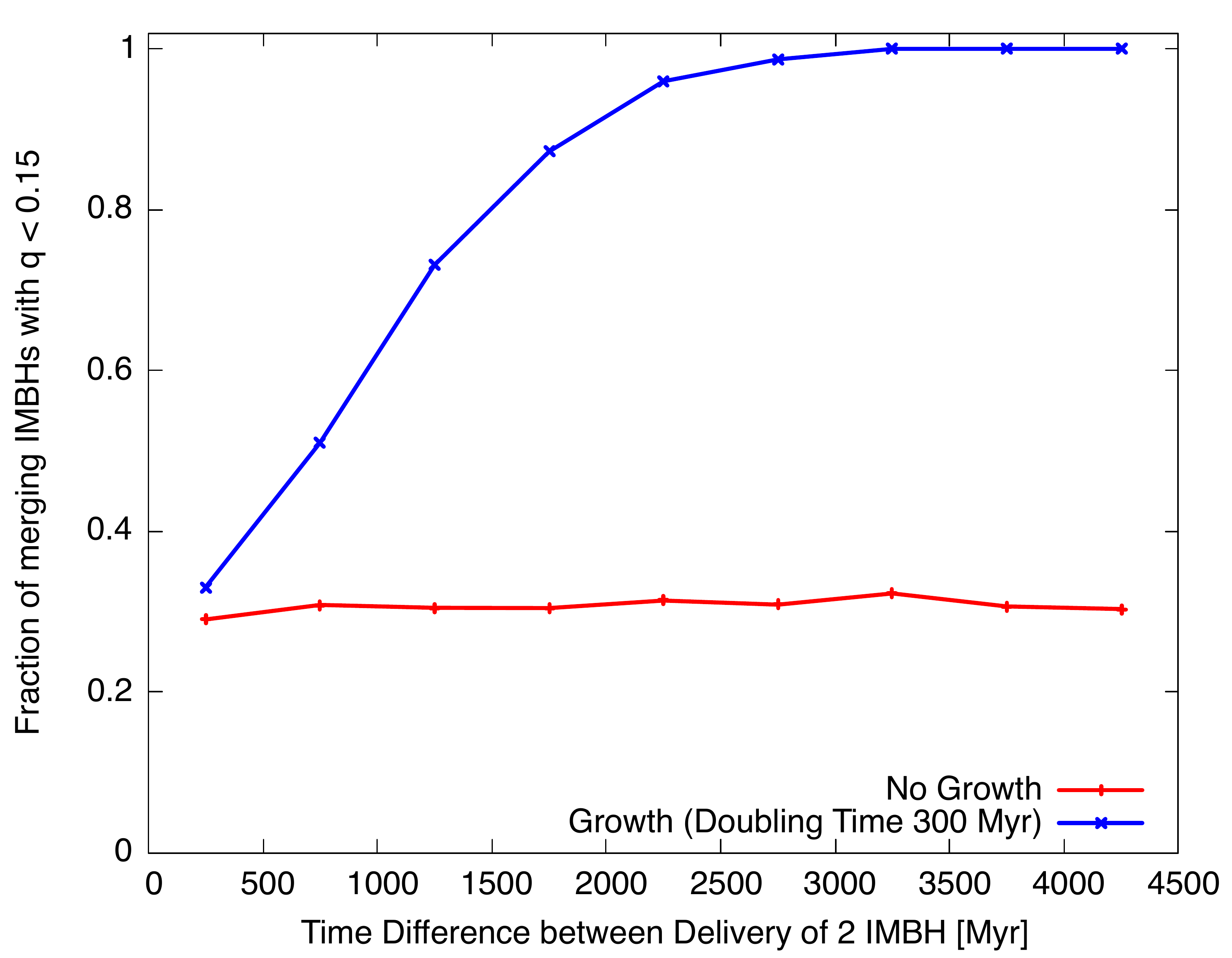}
	\caption{The impact of BH growth on the fraction of binaries in which the mass ratio is less than 0.15 as a function of the time difference between the delivery of the two IMBH. The red line shows the case when no BH growth is considered and the blue line is for the case when the first BH is allowed to grow with a doubling time of 300 Myr.}
    \label{fig:growth-2imbh-case}
\end{figure}

For the \textit{No Growth} case, we find that the seed retention probability for NSC masses between $10^{6.5} -10^{7.5} \ \msun$ is about 50 per cent. It can be seen from Figs. \ref{fig:histo-processed} and \ref{fig:heatmap-processed} that allowing for BH growth with a doubling time of 300 Myr increases this retention probability to about 70 per cent in the same mass range. Allowing for the first BH delivered to grow before the next one is delivered significantly alters the mass ratio distribution of the merging IMBH binary in the case where more than one IMBH are delivered to the NSC. In Fig. \ref{fig:growth-2imbh-case}, we show the fraction of binary IMBHs for which the mass ratio is less than 0.15 as a function of the time difference between the delivery of two IMBH to the NSC. As stated before, it is assumed that the merged BHs will be retained in the NSC if the mass ratio is less than 0.15. The red line is for the case where BH growth is not considered and the blue line is for the case where the BH is allowed to grow with a doubling time of 300 Myr. We see that as the time difference between the delivery of the two IMBH increases, the fraction of binaries with mass ratio less than 0.15 increases sharply since the first BH can significantly increase its mass. Without growth only about 30 per cent of the merged BHs would be retained in the NSC and about 70 per cent would be ejected. Allowing for growth with a doubling time of 300 Myr increases the retention rate to 70 per cent. Allowing for growth at Eddington rates shows an extremely high seed retention fraction of close to 80 to 90 percent for high-mass NSCs. Therefore, allowing for growth has important consequences on the retention of an SMBH seed and the occupation fraction of SMBH for galaxies of different stellar mass.

\begin{figure}
	\includegraphics[width=\columnwidth]{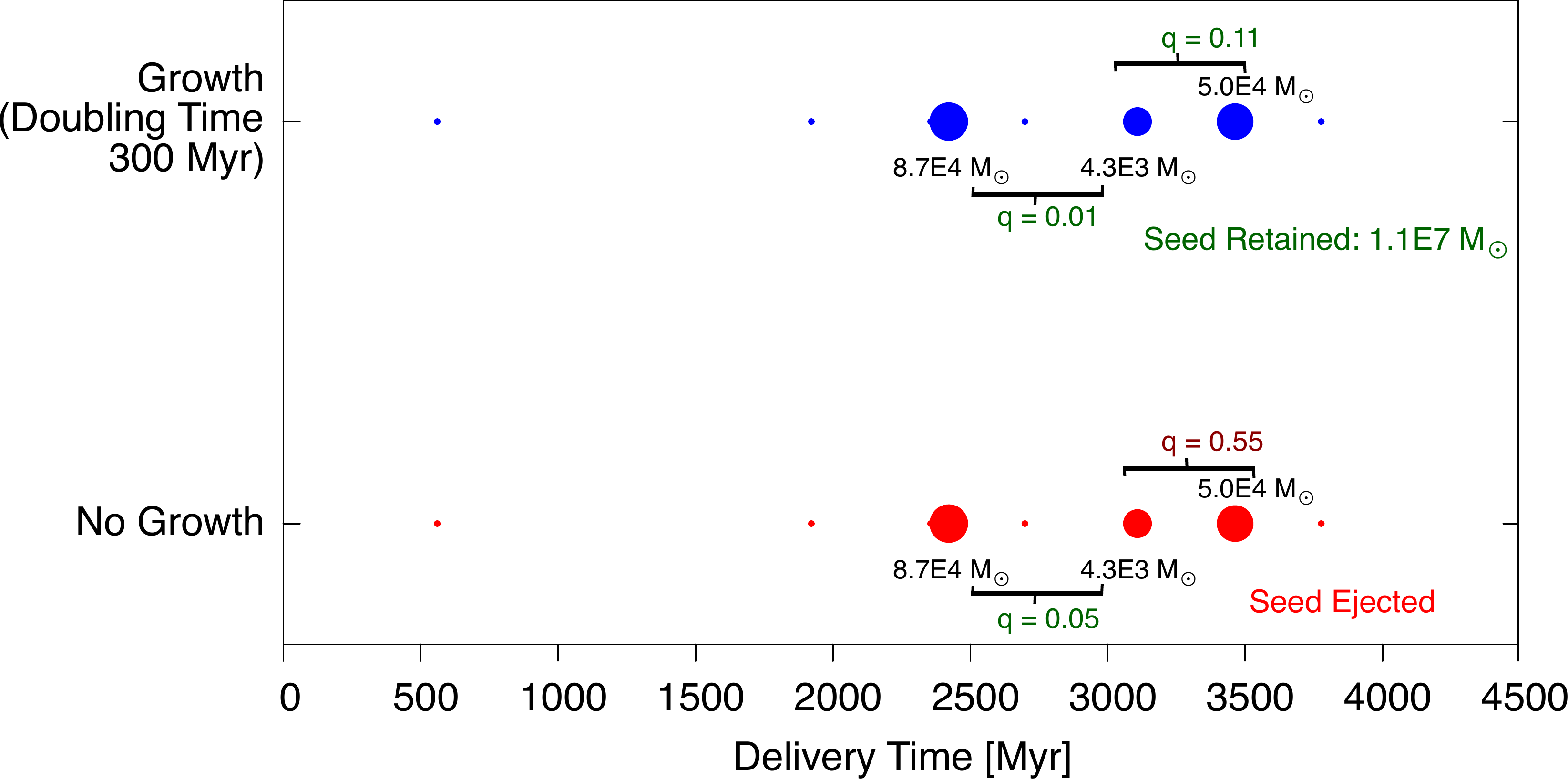}
	\caption{The delivery time of clusters in an example system that forms an NSC with a mass comparable to the Milky Way NSC. For this particular case, eight clusters were aggregated to form the NSC, three of the clusters brought along IMBHs. The first IMBH was delivered at 2421 Myr, the second at 3109 Myr and the last one was delivered at 3465.0 Myr. The red points show the case where we do not allow BHs to grow before merging with the next pair. The merger product of the first IMBH pair is retained since the mass ratio is less than 0.15 but in the next merger the IMBH is ejected due to GW recoil because the mass ratio is larger than 0.15. For the case where we allow growth with a doubling time of 300 Myr, the first IMBH grows sufficiently to have a low-mass ratio when it merges with the other IMBHs,  resulting in the retention of the seed.}
    \label{fig:comparison-growth-no-growth}
\end{figure}

The consequences of allowing for growth are also illustrated in Fig. \ref{fig:comparison-growth-no-growth} where we show the delivery time of clusters in an example model that forms an NSC with mass similar to that of the Milky Way NSC. For this case, eight stellar clusters were aggregated to form the NSC and three of the clusters delivered an IMBH to the NSC. Without accounting for growth, when the first two IMBHs merge, the mass ratio is 0.05 and the merged BH is retained in the NSC. However, when this BH merges with the third IMBH, the mass ratio is 0.55 and therefore the merged BH is ejected from the NSC. However, if we allow for growth with a doubling time of 300 Myr, the mass ratio between the merged IMBH and the third IMBH is 0.11. Therefore, the kick it will receive will be lower and the seed is retained and grows to reach a mass of $1.1 \times 10^{7} \ \msun$.

\section{Discussion and comparison with observations}\label{sec:discussion}

\begin{figure}
	  \includegraphics[width=0.99\linewidth,scale=1.1]{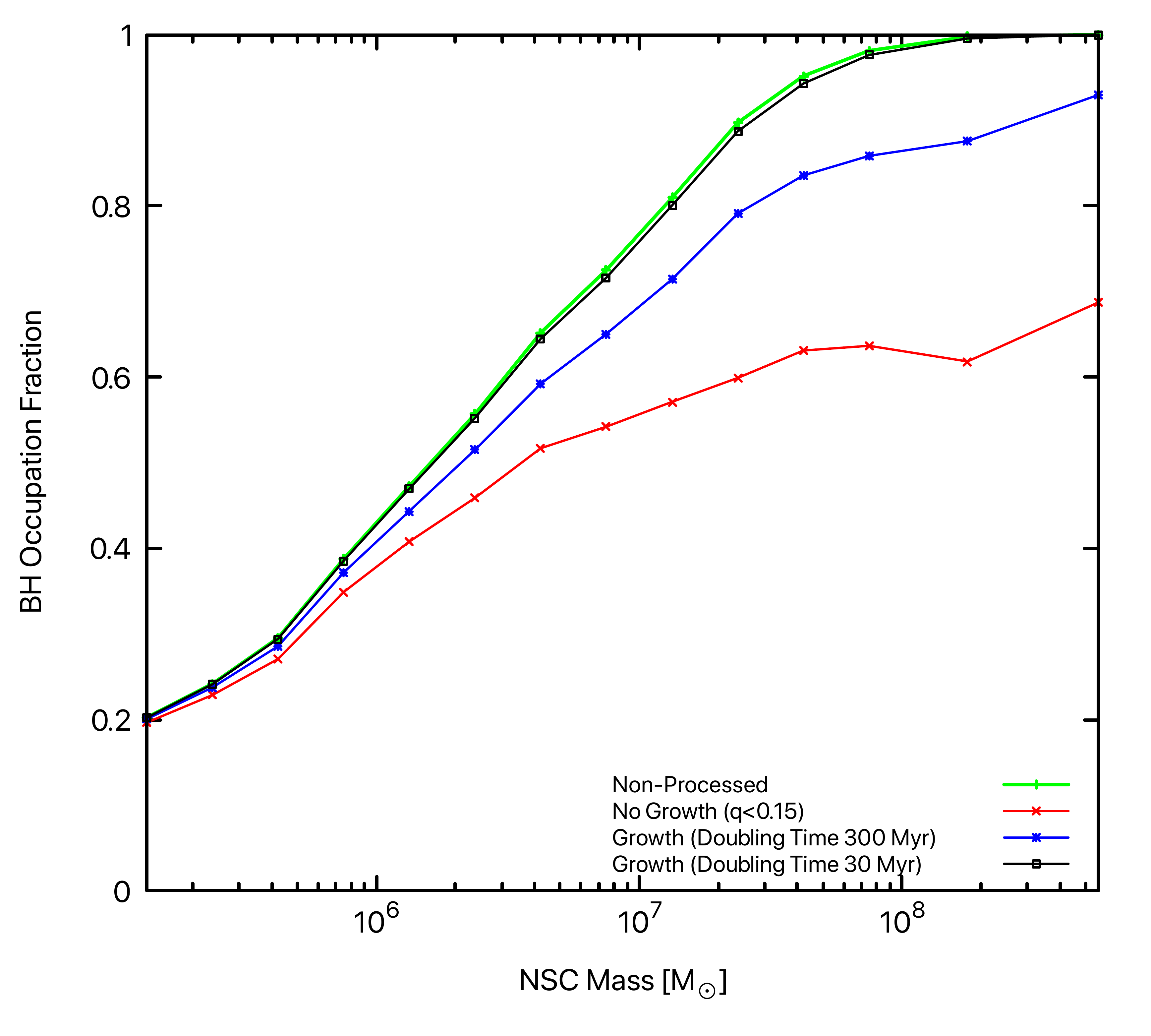}
	\caption{The occupation fraction of seed BHs in an NSC, as a function of NSC mass after accounting for IMBH delivery and possible ejection by GW recoil kicks.
The green line (non-processed) gives the fraction of NSCs in which at least one IMBH was delivered during the assembly of the NSC. The red line shows the occupation fraction of a retained BH seed in the case where we do not consider BH growth. The blue and black lines represent the case where BH growth is considered with a BH mass doubling time of 300 Myr and 30 Myr respectively.}
    \label{fig:occupation-frac-nsc-mass}
\end{figure}

The results presented in Section \ref{sec:results} show that NSCs with masses less than $10^{6} \ \msun$ are highly unlikely to contain an IMBH that could potentially be a seed for the SMBH. However, as NSC mass increases, the probability of stellar clusters bringing along one or more IMBH to the galactic centre increases. In Fig.  \ref{fig:occupation-frac-nsc-mass}, the occupation fraction of an SMBH seed as a function of NSC mass is shown for four different cases. The green line shows the non-processed case, i.e. where, regardless of retention, one or more IMBHs were delivered to the NSC. The red line is for the case where we do not consider BH growth. The blue line and black lines show the cases where we allow for the BH to grow with a doubling time of 300 Myr and 30 Myr respectively.

For NSC mass between $\rm 10^{6} \ \msun$ to $\rm 10^{7} \ \msun$, the BH occupation fraction increases from about 0.2 to 0.5 for the case where we do not allow for BH growth and it increases up to 0.7 when BH growth with a doubling time of 300 Myr is considered. The differences between the left and right panel in Fig. \ref{fig:occupation-frac-nsc-mass} are most prominent at the low NSC mass end due to the inclusion of a probability for forming IMBHs in stellar clusters with a mass less than $10^{5} \ \msun$. SMBH seed occupation fraction increases to up to 0.6 and 0.8 for NSC masses of about $\rm 10^{8} \ \msun$.

\begin{figure}
\includegraphics[width=0.99\linewidth,scale=1.1]{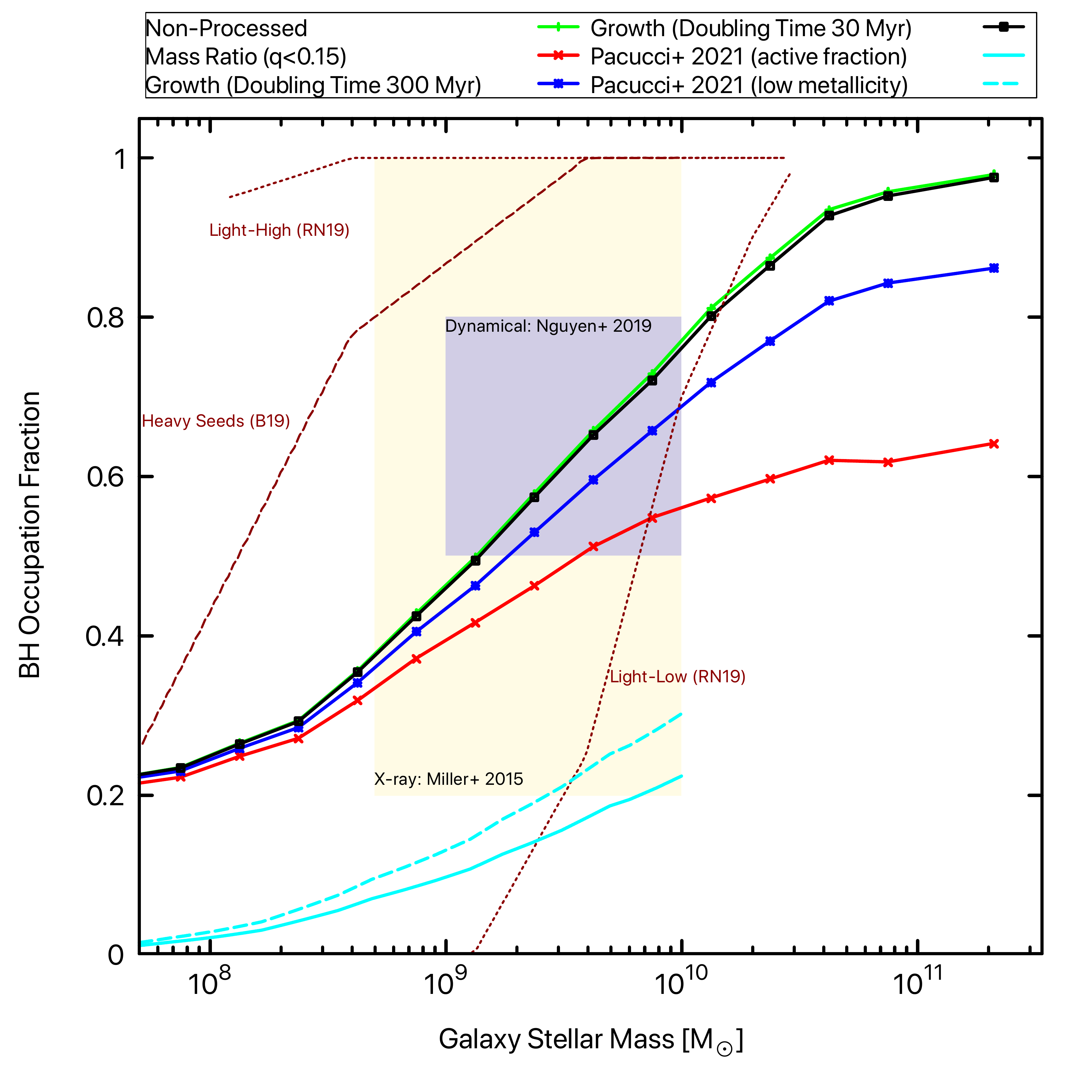}
	\caption{The occupation fraction of BHs as a function of galaxy stellar mass after accounting for IMBH retention in nucleated galaxies.
The lines follow the same colour scheme as Fig. \ref{fig:occupation-frac-nsc-mass}. Observational constraints of BH occupation fraction were taken from Fig. 5 in \citet{greene2019rev}. Constraints on BH occupation fraction from X-ray observations of galactic nuclei are shown in the yellow box \citep{miller2015}. The blue boxes are constraints on the BH occupation fraction from dynamical mass measurements \citep{Nguyen2019}. The dotted dark red lines show the predictions for SMBH ($\rm M> 3 \times 10^{5} \ \msun)$) from the fuelling mechanism of population III seeds by \citet{ricarte18}. The dashed dark red line shows the predicted occupation fraction from \citet{bellovary2019} for BHs assuming heavy initial seeds. The solid and dashed cyan lines show the predicted active fraction of massive BHs in dwarf galaxies based on the theoretical model given in \citet{pacucci2021a}.}
    \label{fig:bh-occupation-frac-galaxy-stellar mass}
\end{figure}

\begin{figure}
	  \includegraphics[width=0.99\linewidth,scale=1.1]{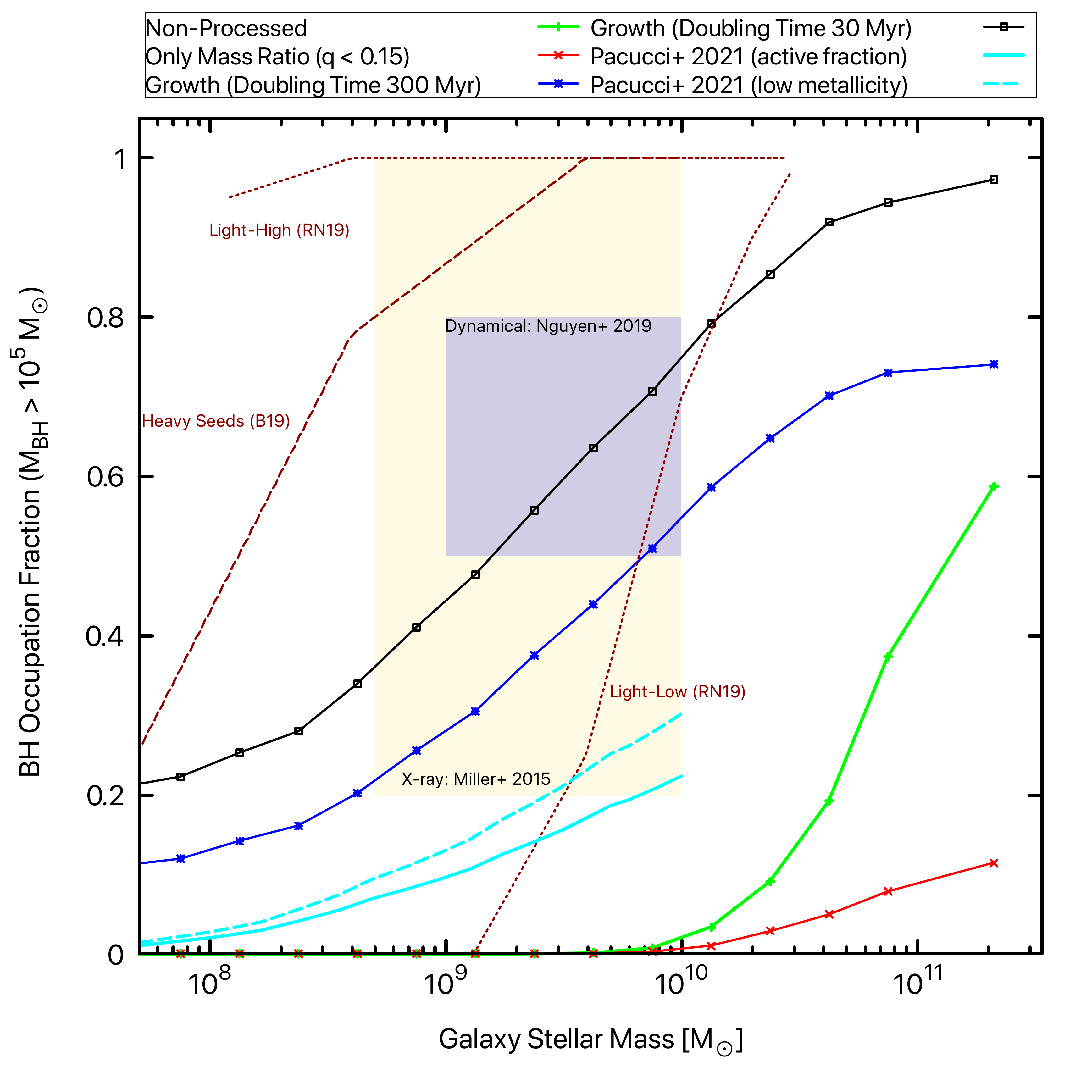}
	\caption{The occupation fraction of BHs more massive than $10^{5}\,\msun$ as a function of galaxy stellar mass after accounting IMBH retention.
If BH growth is not considered (red line), the occupation fraction of BHs more massive than massive than $10^{5}\,\msun$ is close to zero up to about $10^{10}\,\msun$ stellar mass galaxy. This is because the majority of the IMBH delivered to the NSC by stellar clusters are between $10^{2}$ to $10^{4}\,\msun$ However, allowing for BH growth (blue and black lines) significantly increases the occupation fraction of BHs more than $10^{5}\,\msun$ with galaxy stellar mass.}
    \label{fig:massive-bh-occupation-frac-galaxy-stellar mass}
\end{figure}

Since the NSC mass derived in our population synthesis study depends on the galaxy stellar mass. We can show the BH occupation fraction in nucleated galaxies as a function of galaxy stellar mass. This is done in Fig. \ref{fig:bh-occupation-frac-galaxy-stellar mass}. We have also added observational constraints on BH occupation fraction taken from Fig. 5 in \citet{greene2019rev}. It can be seen from the Fig. \ref{fig:bh-occupation-frac-galaxy-stellar mass} that the occupation fraction for massive BHs predicted from our seeding mechanism is within the observational constraints from dynamical and X-ray observations. The seeding mechanism proposed here also places strong constraints on the occupation fraction of massive BHs in galaxies lower than $10^{9} \ \msun$ where the predicted occupation fraction is between 0.2 to 0.3. In Fig. \ref{fig:massive-bh-occupation-frac-galaxy-stellar mass}, the occupation fraction of BHs more massive than $10^{5}\,\msun$ is shown. Given the seed masses we start out with, the occupation fraction for these massive BHs is close to zero for galaxies with stellar mass less than $10^{10} \ \msun$ for the case where we do not allow BH growth. However, if BHs are allowed to grow with a mass doubling time of 300 Myr, the occupation fraction of these massive BHs increases significantly (blue line in Fig. \ref{fig:massive-bh-occupation-frac-galaxy-stellar mass}). 

\begin{figure}
\includegraphics[width=\columnwidth]{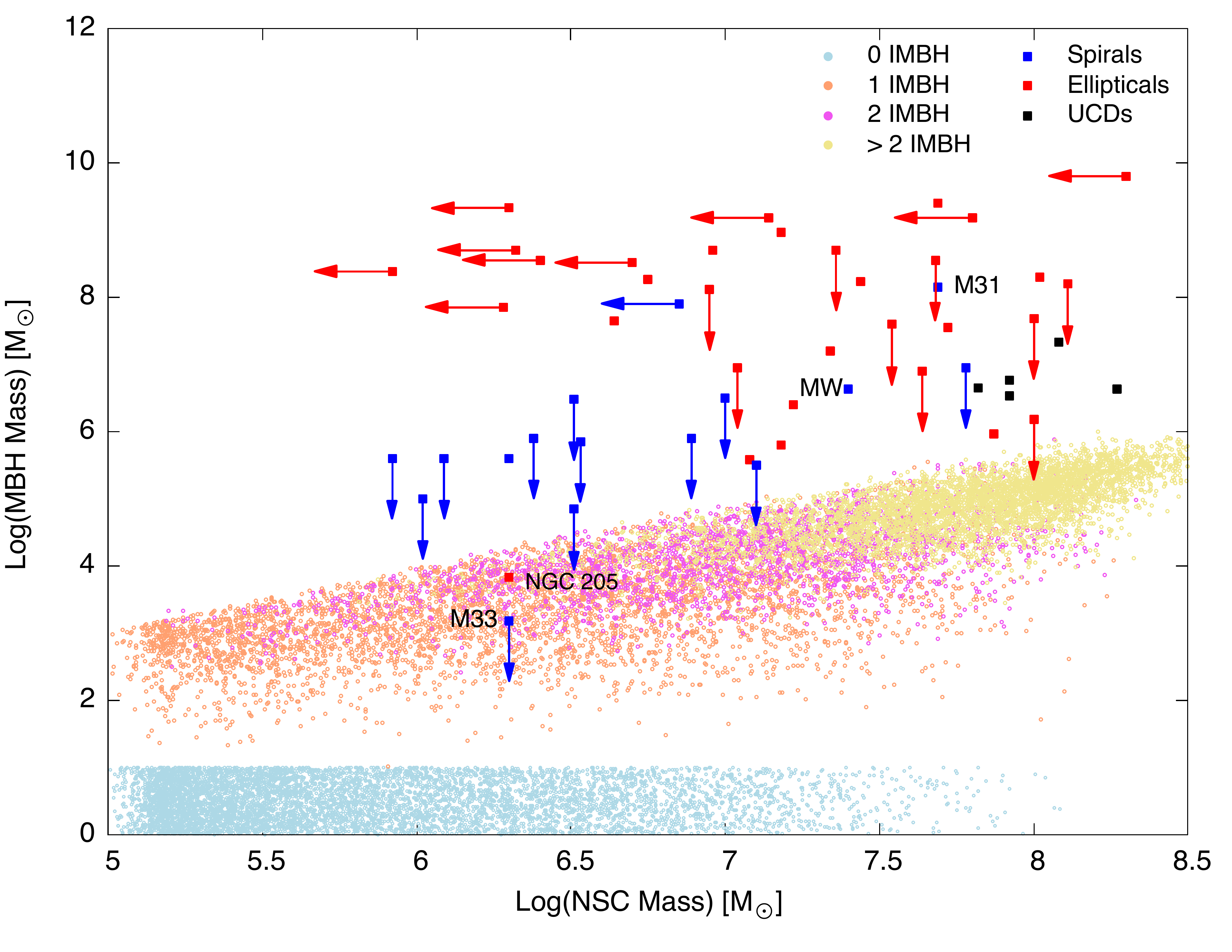}
	\caption{Observed galaxies for which we have constraints on NSC Mass (x-axis) and SMBH Mass (y-axis) shown in blue (spiral galaxies), red (elliptical galaxies) and black points (ultracompact dwarfs). The blue points are our NSC models in which no IMBHs are delivered, they have been artificially spread on the y-axis between zero and $10 \ \msun$. The orange data points are model galaxies where one IMBH is delivered to the NSC, the pink points are where two IMBHs were delivered to the NSC and yellow points are NSCs where more than two IMBHs were delivered. This data assumes low BH natal kicks in simulated GC models (see probability values in Table \ref{tab:reaslitic}). To avoid saturation in the plot, we only show a limited number of points from our population synthesis results. Several hundred representative points were randomly picked in different NSC mass bins.}
    \label{fig:nsc=mass-vs-bhmass-obs-sim}
\end{figure}

\begin{figure}
\includegraphics[width=\columnwidth]{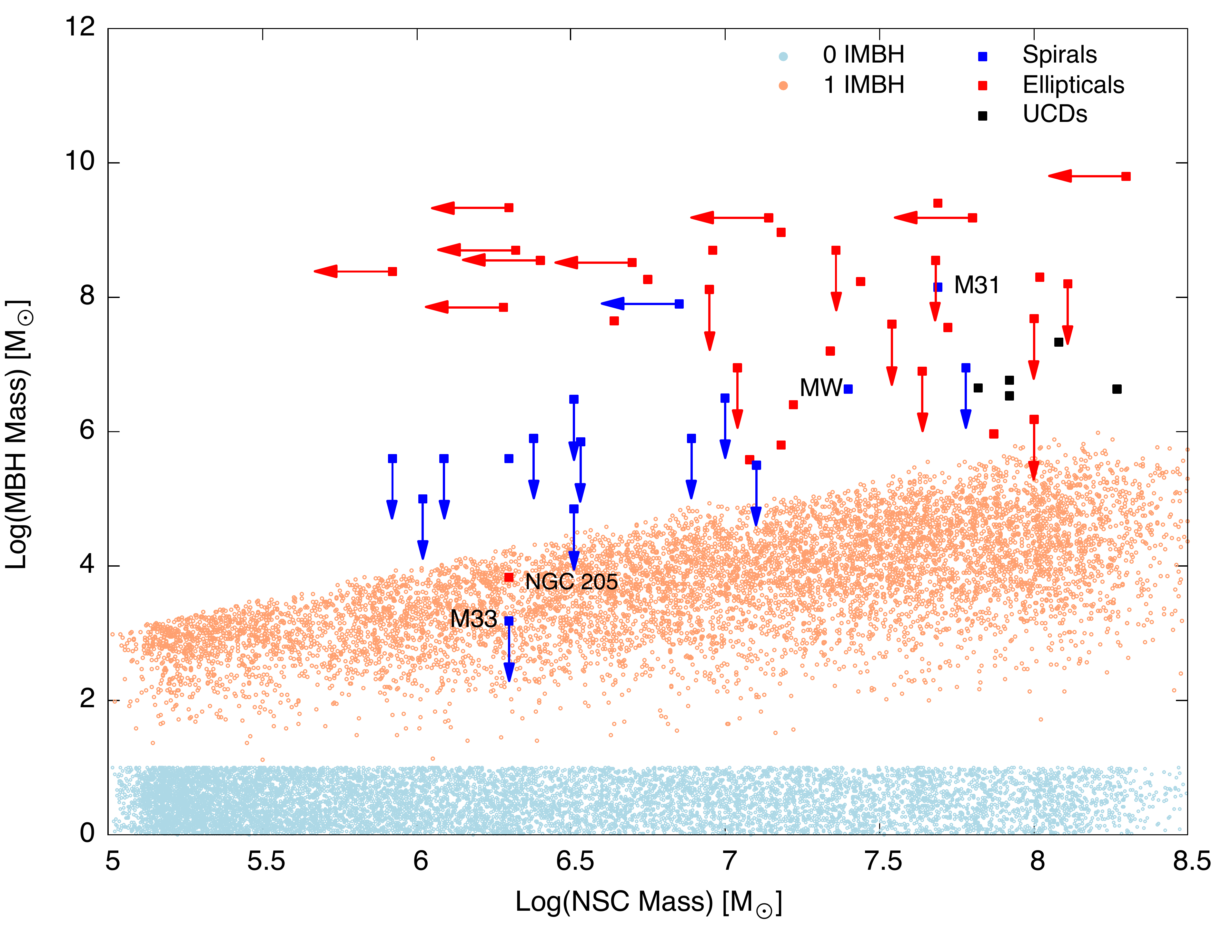}
	\caption{Results from our population synthesis approach for the case where we have no BH growth. Retention of merged IMBH pairs is only allowed if their mass ratio (q) is $\le$ 0.15. At the low NSC mass end, there is a high likelihood that either zero or one BH would be delivered to the NSC (see Fig. \ref{fig:histo-delivered-low}). For NSCs where two IMBHs are delivered, the likelihood of retaining a seed BH is low. The BH masses that we sample depend on the mass of the stellar cluster in which they formed and only the most massive NSCs contain BHs that are $10^{4}$ to $10^{5} \ \msun$}
    \label{fig:nsc=mass-vs-bhmass-no-growth}
\end{figure}

\begin{figure}
 \includegraphics[width=\columnwidth]{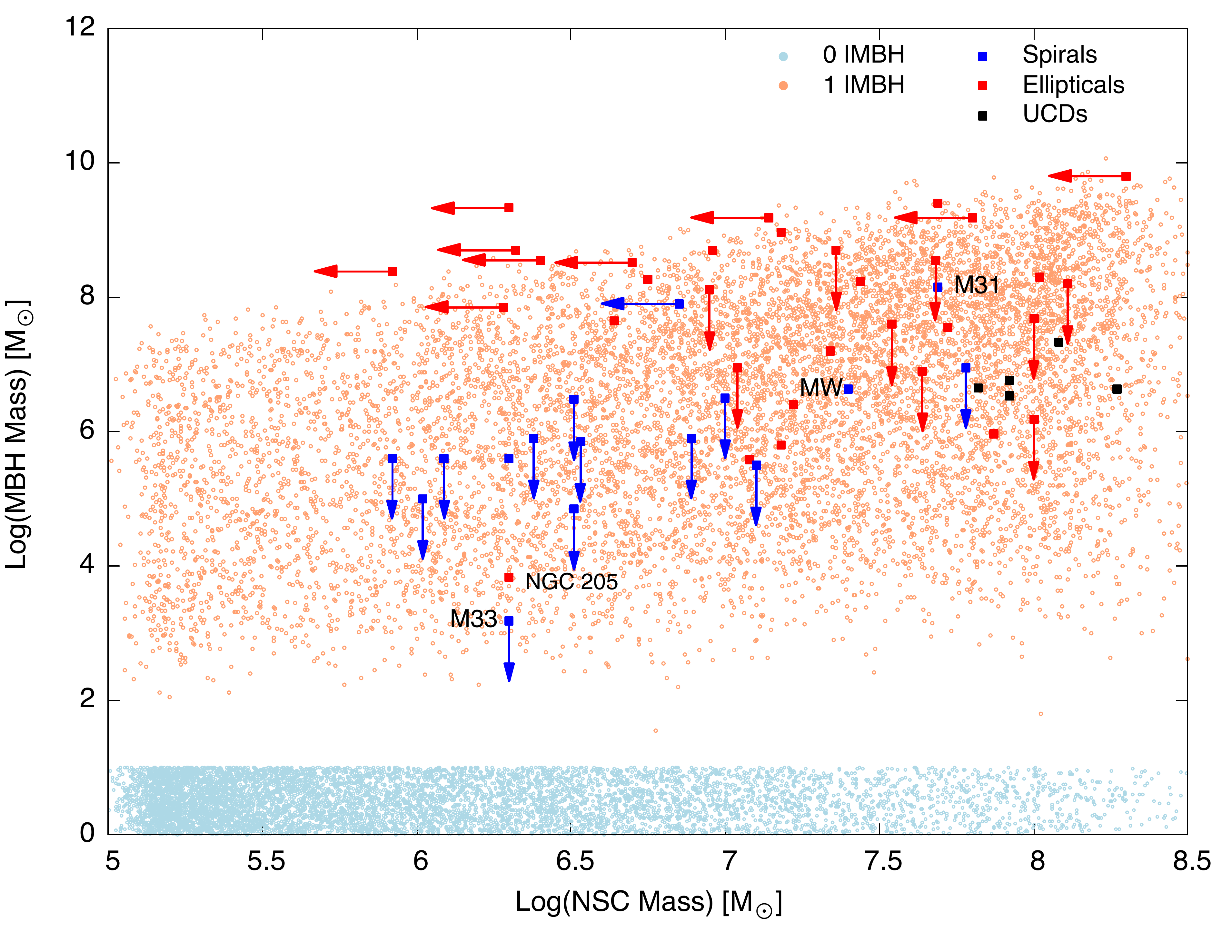}
	\caption{Same as Fig. \ref{fig:nsc=mass-vs-bhmass-no-growth}. However, in this case we allow for BHs to grow with a mass doubling time of 300 Myr. The spread in the BH masses arises from their delivery time to the NSC which was randomly sampled between 0 and 4500 Myr. Seed BHs that are delivered later on have less time to grow in the NSC.}
    \label{fig:nsc=mass-vs-bhmass-growth300myr}
\end{figure}

In Figure \ref{fig:obs-nsc-vs-bh}, we showed the NSC mass vs BH mass for observed galaxies with constraints on both values \citep{Neumayer2012,scott2013,Graham2016,Nguyen2018,Nguyen2019}. We now compare how the NSC masses and BH masses generated from our population synthesis approach compare with these observations. In Fig.  \ref{fig:nsc=mass-vs-bhmass-obs-sim}, we overplot our sampled IMBH masses for galaxies with NSCs that contain multiple IMBHs with orange (one IMBH), pink (two IMBH) and yellow points (more than two IMBH). NSCs that do not contain an IMBH are shown with the light blue points which have been artificially spread between zero and one. For visibility, a smaller representative subset of the sampled data is plotted in \ref{fig:nsc=mass-vs-bhmass-obs-sim}. In this figure, we do not account for the growth of BHs and simply show the sum of the masses of the IMBH that are delivered. The most massive NSCs ($10^{7.5}-10^{8.5} \ \msun$) can contain seeds IMBHs close to $10^{5} \ \msun$. These NSCs are produced from the merger of many stellar clusters, which then increases the total mass of delivered IMBHs. Additionally, some of these NSCs can be constructed from massive stellar clusters that are likely to form a more massive IMBH.

In Fig. \ref{fig:nsc=mass-vs-bhmass-no-growth}, we show the results where we either merge and retain or merge and eject multiple seed IMBH without BH growth. It can be seen from the figure that we are unable to reproduce observed constraints and upper mass limits on observed BH masses. The growth of IMBHs through mergers with each other is not enough to account for observations of well-observed SMBHs like those in the Milky Way and M31. In order to reproduce observed SMBH masses in galaxies with NSC that are more massive than $3 \times 10^{6} \ \msun$, we need to grow the IMBH by one to three orders of magnitude. For less massive NSCs, e.g., NGC 205, the total IMBH mass is consistent with observational upper limits of $10^{5}\ \msun$ for the SMBH.

\begin{figure}
	 \includegraphics[angle=0,width=\columnwidth]{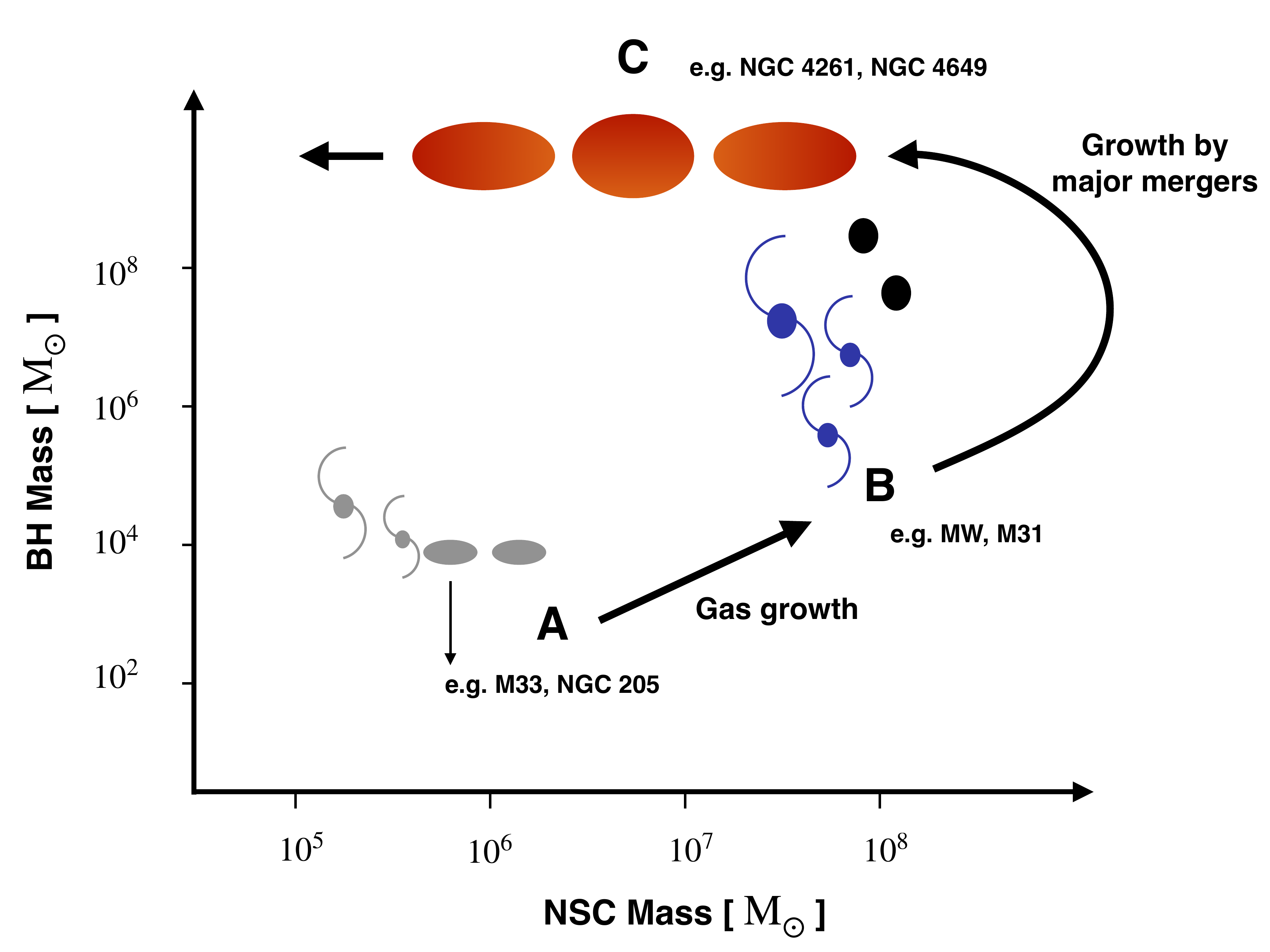}
	\caption{Schematic illustration showing different classifications of observed galaxies that have measurements or constraints on NSC and BH masses (see Fig. \ref{fig:obs-nsc-vs-bh}). ``A'' type galaxies are low-mass spiral and dwarf elliptical galaxies, e.g., M33, NGC 205. These galaxies have relatively low NSC masses (less than few $10^{6} \ \msun$) and upper limits on BH masses are between $10^{3}$ to $10^{6} \ \msun$. Our population synthesis results indicate the NSC in these galaxies can form from the merger of few stellar clusters and that it is highly likely that none or one seed BH is delivered to the NSC (see Section \ref{subsec:a-type}). ``B'' type galaxies have NSC masses larger than several $10^{6} \ \msun$ and their SMBH masses are between $10^{6} \ \msun$ to $10^{8} \ \msun$. Growth due to the presence of gas in the nuclei of these galaxies could account for their more massive SMBH and may also contribute to the growth of the NSC through in-situ star formation (see Section \ref{subsec:b-type}). ``C'' type galaxies are massive ellipticals that contain an SMBH but have a low-mass NSC. The processes which could lead to decreased NSC mass in these galaxies are described in Section \ref{subsec:c-type}}.
    \label{fig:illustration}
\end{figure}

Fig. \ref{fig:nsc=mass-vs-bhmass-growth300myr} shows the distribution of retained SMBH seeds on the same figure for the case where we allow the BH to grow with a mass doubling time of 300 Myr. With this assumption, we not only get a higher occupation fraction of an SMBH seed with increasing NSC mass, but we are also able to account for the observed SMBH masses. This assumption also predicts a larger number of massive BHs for galaxies with low NSC masses (few $10^{6} \ \msun$). Our results suggest that such NSCs are most likely to contain at most one IMBH. The growth of the seed BHs in these lower-mass galaxies will depend on how much gas there is in them. Fig. \ref{fig:nsc=mass-vs-bhmass-growth300myr} shows that we are probably over-growing BH masses for low-mass galaxies with NSC masses between $10^{6}$ to $10^{6.5} \ \msun$. It is likely that these galaxies did not capture enough gas to grow their central BHs at 10 per cent of the Eddington rate for an extended period of time. Absence of gas in these galaxies could also account for their low galaxy stellar mass, NSC mass and BH mass. In contrast, it is also likely that galaxies with NSC mass larger than $\rm 10^{7} \ \msun$ contain more gas that would allow the BH seed to grow via accretion. This gas may also contribute to in-situ star formation in the NSC leading to its growth. Therefore, it is likely that in galaxies with stellar masses larger than few $10^{9} \ \msun$, in-situ growth of the NSC from accreted gas dominates the growth of the NSC and SMBH seed (See Fig. \ref{fig:illustration}).

In the next few subsections, we specifically discuss the formation and evolution of the galactic nuclei in galaxies categorized as ``A'' type, ``B'' type and ``C'' type in Fig. \ref{fig:obs-nsc-vs-bh}.

\subsubsection{``A'' type galaxies}\label{subsec:a-type}

M33 is a relatively low-mass local group spiral galaxy with an NSC of around $\sim \ 2 \times 10^{6} \ \msun$ \citep{kormendy1993,graham2009}. However, it does not contain an SMBH and upper limits for a central BH mass is about $1500 \ \msun$ \citep{Merritt2001,Gebhardt2001}. Similarly,
the dwarf elliptical galaxy NGC 205 has an NSC with mass $\sim \ 2 \times 10^{6} \ \msun$ and recent observations indicate a presence of $6760 \ \msun$ central BH \citep{Nguyen2019}. Our results indicate that for low-mass galaxies with NSC mass between $1-3.2 \times 10^{6} \ \msun$, the probability of having no SMBH seed delivered to the NSC is about 50 percent. So it is very likely that galaxies such as M33 and NGC 205 had either:
\begin{enumerate}
    \item Never had an SMBH seed to begin with, because none of the clusters that merged to form the NSC had formed an IMBH
    ($\sim$ 50 per cent probability from our analysis).
    \item Had a single BH seed that was not very massive ($10^{3}$ to $10^{4} \ \msun$) and did not grow substantially due to limited gas accretion, e.g., in a dwarf elliptical galaxy ($\sim$ 40 per cent probability from our analysis). This channel may also explain observed central BH masses between $10^{3}$ to $10^{4}\ \msun$ in galaxies with stellar masses similar to M33, like NGC 4395 \citep{woo2019,cho2021}, NGC 3319 \citep{jiang2018,davisgraham2021} and NGC 205 \citep{Nguyen2019}.
    \item  Had two IMBHs delivered to the NSC ($\sim$ 10 per cent probability from our analysis). If these IMBHs promptly merged due to GW radiation (as seen in runs 2.2 and 2.3 of \citetalias{askar2021}), then depending on the mass ratio of the merging BHs, the merger product could potentially be ejected from the NSC due to the GW recoil kick. Such ejected BHs may still be bound to the galaxy and may contribute to the population of wandering massive BHs \citep{reines2020,guo2020,ricarte2021,greene2021}.
\end{enumerate}

\subsubsection{``B'' type galaxies}\label{subsec:b-type}

``B'' type galaxies like the Milky Way and M31 host more massive NSCs (
$10^{7} \ - 10^{8} \msun$) and also harbour SMBHs of $10^{6}$ to few $10^{8} \msun$ \citep{bland-hawthorn2016,williams2017}. We find that for galaxies that host NSCs in the mass range $10^{7} \ - 10^{8} \ \msun$, the likelihood of no IMBHs being delivered to their centres is less than 10 per cent.

For NSC masses in the range $10^{7}$ to $10^{7.5} \ \msun$, the probability of having one or two IMBHs delivered to the centre is about 25 per cent each. The probability of having three IMBHs delivered to the centre is about 20 per cent and the probability of four or more IMBHs being delivered is also about 20 per cent. The following scenarios may occur in galaxies with NSC masses within this range:
\begin{enumerate}

  \item If one IMBH is delivered then it may grow through gas accretion. Allowing for 10 per cent of Eddington accretion on to the BH (within a time of up to 4.5 Gyr), we find that BH masses can be in the range of $10^{3}$ to $10^{8} \ \msun$. In this case, the NSC could also grow by in-situ star formation from the accreted gas that feeds the BH. So NSCs with masses originally in the range $10^{6} \ \msun$ to $10^{7} \ \msun$ could grow in mass by a factor of a few in these galaxies. The probability of having a single IMBH seed delivered in such NSCs is about 40 per cent.
    
  \item If two IMBHs are delivered to the cluster centre then they will likely form an IMBH binary (as shown in the previous paper in this series, \citetalias{askar2021}). If the binary components merge due to GW radiation emission and their mass ratio is small (less than 0.15) then the IMBH seed can be retained in the cluster \citepalias[see Section 6.1 in][]{askar2021}. If we allow for the first IMBH to grow through gas accretion before the second IMBH is delivered then the mass ratio of the merging IMBH will be small. Assuming that the IMBH binary will merge promptly, we find that for NSC masses originally in the range $10^{7}$ to $10^{7.5} \ \msun$, the chance of retaining an SMBH seed will be about 80 per cent. For NSC masses originally in the range $10^{6}$ to $10^{7} \ \msun$ the probability of retaining an SMBH seed will be between 50 to 70 per cent (see Figs. \ref{fig:histo-processed} and \ref{fig:heatmap-processed}).
    
  \item If three or more IMBHs are delivered to the NSC then few-body interactions between the IMBHs will be possible (as seen in runs 3.1 and 3.3 in \citetalias{askar2021}). In these cases, we find the first two IMBHs that are delivered form a binary and typically scatter away the third IMBH. This may increase the merger time for the IMBH binary as its orbital eccentricity can decrease in a few-body interactions involving IMBHs that have similar mass (as seen in run 3.1 of \citet{askar2017}).  This scenario is further discussed in Section \ref{subsec:imbh-merger-interactions}. It is more likely that the third IMBH is delivered after the first two IMBHs have already merged due to GW radiation. Depending on whether the merged BH is retained or ejected, we may either end up with an IMBH binary or a single IMBH in the NSC.

\end{enumerate}

\subsubsection{``C'' type galaxies}\label{subsec:c-type}

``C'' type galaxies are elliptical galaxies that have massive BHs in the range $10^{7}$ to $10^{9} \ \msun$. Their NSC masses are small with upper limits typically between $10^{6}$ to $10^{7} \ \msun$. The observed fraction of nucleation in massive elliptical galaxies (with stellar mass between few $10^{11}$ to $10^{12} \ \msun$) is less than 0.35 (see Fig.\,3 in \citet{Neumayer2020}). In order to explain the presence of a massive BH with a low-mass NSC in these galaxies it is possible that:
\begin{enumerate}
    \item these galaxies are likely formed via major mergers. thus it may be possible to form an SMBH binary in the NSC of these elliptical galaxies (e.g., the recent detection of a dual SMBH system in the irregular galaxy NGC 7727 by \citet{voggel2021}). This binary could merge through GW radiation and in the process of merging it can transfer energy to surrounding stars. This dynamical heating decreases the stellar density in the galactic nuclei and can effectively destroy the NSC \citep{cote2006,gualandris2009,bekki2010,antonini2015}.
    \item the high velocity dispersion in the presence of a massive SMBH will drive up the collision rate of stars and this could deplete stars in the vicinity of the BH \citep{Davies2011,amb2021}.
   \item More massive SMBHs can tidally disrupt infalling stellar clusters that are much further away from them \citep{antonini2013}. This will prevent the formation of a dense cluster of stars close to the BH. Therefore, if an NSC is destroyed in these galaxies then it cannot re-form.
\end{enumerate}

\subsection{Caveats and Limitations}\label{subsec:caveats}

This subsection discusses the main caveats and limitations connected with the assumptions that go into the population synthesis approach used in this paper.

\subsubsection{Ad-hoc model for BH growth}\label{ad-hoc-bh-growth}

The assumption for the BH growth rate through gas accretion most strongly affects the retention fraction of a seed SMBH in more massive NSCs in which two or more IMBHs are likely to be delivered. Most of these NSCs have a mass larger than $\sim 10^{7} \ \msun$ (see Fig. \ref{fig:histo-delivered-low}). In the extreme case, where we considered a BH mass doubling time of 30 Myr (assuming that the BH is accreting at the Eddington limit), we found that the seed SMBH retention fraction is higher than 85 per cent for NSCs more massive than about $10^{7.5} \ \msun$ (see Fig. \ref{fig:heatmap-processed}). In the other extreme case, where we do not consider BH growth, the seed SMBH retention fraction for massive NSCs is only 60 per cent.

Modelling in-situ growth of NSCs and seed BHs from accretion is challenging given the difficulty in accounting for relevant physical processes, feedback and the different timescales involved. Therefore, we consider an ad-hoc model for BH growth to check what consequences that would have on the SMBH seeding mechanism proposed in this paper. One of the main assumptions in this work is that the NSC assembly from merging stellar clusters occurs within an interval of 4500 Myr (between redshifts $1<z<4$). We randomly sample the delivery time of IMBHs in the NSC within this 4500 Myr interval. The delivery time has important consequences for the case where we consider BH growth. If a single IMBH is delivered within the first few hundred Myr then it can significantly increase its mass from the time it was delivered to up to 4500 Myr (up to about 15 doubling times assuming that the BH is accreting at an average rate of 10 per cent of the Eddington rate over the time it is allowed to grow in the NSC, i.e., the doubling time is 300 Myr). For NSCs with multiple IMBHs, the delivery time difference between the two IMBHs has important consequences on the GW recoil kick and subsequent retention of the merged BH (see Fig. \ref{fig:growth-2imbh-case}). Average growth history of SMBHs computed from observed X-ray active galactic nuclei (AGN) luminosity functions \citep{soltan1982, marconi2004} suggest that most SMBHs underwent significant growth between redshifts 3 and 1 \citep[see Fig. 2 in][]{brandt2010}. These results are also corroborated by simulations of galaxy formation and evolution that consider AGN growth \citep{dimatteo2008,fanidakis2012}. Therefore assuming that the BH grows at about 10 per cent of the Eddington accretion rate over the period that we consider is reasonable and allows us to reproduce observed SMBH masses (as shown in Fig. \ref{fig:nsc=mass-vs-bhmass-growth300myr}). \citet{pacucci2021a} developed a theoretical model that uses the properties of the host galaxy to predict what fraction of them could contain an active massive BH. They find that the active fraction of accreting SMBHs ranges between 5 to 22 per cent depending on the host galaxy stellar mass. In Fig \ref{fig:bh-occupation-frac-galaxy-stellar mass}, we show the predicted occupation fraction from \citet{pacucci2021a} with cyan lines. If 10 per cent of the BHs in dwarf galaxies are active then dividing our occupation fraction (shown with dark blue line in Fig. \ref{fig:bh-occupation-frac-galaxy-stellar mass}) by 10 gives roughly the same values as those predicted by \citet{pacucci2021a}.

AGN observations show that they have diverse values for Eddington ratios and that mean values are close to 0.1, albeit with significant dispersion \citep{shen2014,panda2018}. The seeding mechanism proposed in this paper may not be sufficient for very high-redshift AGNs. However, a lower-mass seed BH delivered by stellar clusters may grow rapidly during early episodes of chaotic accretion \citep{king2008,alexander2014,zubovas2021,das2021} and hyper-Eddington accretion \citep{pacucci2015, inyoshi2016} in gas rich nuclei. These processes may lead to shorter timescales for BH growth in the galactic nuclei. Significant growth of an SMBH seed may then lead to higher retention rates due to lower GW recoil kicks following mergers with lower mass BHs. Therefore, the occupation fraction of massive BHs in gas-rich environments may be expected to be larger than what we predict with our scenario. 

Additionally, BH growth inside the NSC through gas accretion also implies that a significant reservoir of gas, which could lead to the in-situ formation of stars inside an NSC \citep{antonini2015,guillard2016,AMB2019b}. This can increase the NSC mass that we calculate from the dry merger of stellar clusters. This in-situ growth of NSCs will become important in more massive galaxies (``B'' type galaxies in Figs. \ref{fig:obs-nsc-vs-bh} and Figs. \ref{fig:illustration}). For low-mass galaxies with NSC less massive than few $10^{6} \ \msun$ (``A'' type galaxies in Figs. \ref{fig:obs-nsc-vs-bh} and Figs. \ref{fig:illustration}), we may be over estimating the BH masses since these galaxies may not have had significant gas that can either be accreted by the BH or that may result in the in-situ growth of the NSC.

\subsubsection{IMBH formation and growth in stellar clusters}\label{subsec-caveats:imbh-formation-growth}

The other major assumption that goes into our work concerns the formation time and masses of IMBH that form in stellar clusters. The IMBH formation probability and mass estimates (see Tables \ref{tab:reaslitic} and \ref{tab:sample-imbh-mass}) that we used were based on results from hundreds of Monte Carlo simulations of stellar clusters with different initial parameters carried out using the \textsc{mocca} code. As discussed in \citetalias{askar2021}, in a significant fraction of these simulated stellar clusters, an IMBH forms within tens of Myr due to runaway mergers between massive main-sequence stars \citep{SPZ02,Giersz15,dicarlo2021} leading to the formation of a massive star which may evolve into an IMBH in low metallicity clusters. It is also possible for these stars to absorb a stellar-mass BH \citep{rizutto2020}. In the latter case, the simulated \textsc{mocca} models assumed that all of the mass of the star is transferred to the BH in such mergers. This assumption facilitates the formation and growth of an IMBH seed. It was shown in \citet{Giersz15} that an IMBH may still form in dense clusters through such mergers if the BH only accretes about 25 per cent of the mass of the star. Additionally, if IMBH formation occurs from the mergers of stellar-mass BHs, then depending on the mass ratio, spin magnitudes and directions of the merging BHs, the GW recoil kick may eject the IMBH from the stellar cluster \citep{oleary2006,holley2008,gerosa2021b}. \citet{morawski2018} showed that accounting for these GW recoil kicks may inhibit IMBH formation in \textsc{mocca} simulations in up to 70 per cent of the cases. Taking this into account, we scaled the probabilities for IMBH formation in clusters with a given initial mass (see Fig \ref{fig:gc-mass-imbh} and Table \ref{tab:reaslitic}).

The possible formation and growth of IMBHs in stellar clusters remain uncertain and it is likely that it strongly depends on the initial density, mass, binary fraction and metallicity of the cluster \citep{Sedda2019,hong2020}. The search for IMBHs in Galactic globular clusters using radio observations have not found conclusive evidence for their presence \citep{tremou18} and the upper mass limit for IMBHs in these globular clusters is about $1000 \ \msun$ \citep{baumgardt2017,greene2019rev}. However, recent observational results from extragalactic stellar clusters indicate that some of them may harbour IMBHs. For instance, \citet{lin18} discovered an IMBH candidate in an extragalactic stellar cluster by observing an X-ray flare that originated from a tidal disruption event \citep{lin2020,wen2021}. Several IMBH candidates have also been identified in low-luminosity AGNs \citep{greene2007,baldassare2015,chilingarian2018}. More recently, \citet{Nguyen2019} used dynamical mass measurements to constrain the mass of a BH at the centre of NGC 205 to be less than $7 \times 10^{4} \, \msun$. Additionally, \citet{pechetti2021} detected a $10^{5} \, \msun$ IMBH in the most massive stellar cluster in M31. This cluster is likely to be a stripped NSC of a $10^{9} \ \msun$ galaxy. These observations suggest that a fraction of stellar clusters may have formed an IMBH and it is likely that these stellar clusters were initially massive and dense. Since the strength of the dynamical friction which a cluster undergoes depends linearly on its mass, a significant fraction of such clusters may have inspiraled and merged with the nuclei of their host galaxy \citep{Sedda2019}.

We assume in our population synthesis study that about 25 to 30 per cent of the most massive stellar clusters (>$10^{6} \ \msun$) can form an IMBH which has a mass which is at the most one per cent of the initial cluster mass. If we decrease the probability for IMBH formation in a given stellar cluster, then the number of IMBHs delivered to the NSC would be lower and this can decrease the occupation fraction of seed SMBHs. It can be seen from the results shown in Figs. \ref{fig:histo-delivered-low} and \ref{fig:heatmap-all} that for most low-mass NSCs, the likelihood of delivering none or one IMBH is the highest. This is because only a few stellar clusters need to be aggregated to form the NSC. For high NSC masses (few $10^{7} \ \msun$), the delivery of multiple IMBHs becomes more probable since many more stellar clusters are needed to form the NSC. Also the likelihood of having more massive stellar clusters merge to form the NSC increases. Therefore, even with a decreased probability of IMBH formation in a given stellar cluster, it is possible for one of the many merging stellar clusters to deliver an IMBH to the NSC. Due to this, we can expect that the SMBH seed occupation fraction might not change significantly for these high-mass NSCs. 


Our prescription for estimating the IMBH mass in our population synthesis results gives a mean IMBH mass of about $5 \times 10^{3} \ \msun$ and a median mass of about $1.5 \times 10^{3} \ \msun$. These estimated masses are consistent with constraints derived from simulations of young massive clusters and semi-analytical estimates for BH growth in dense environments \citep{antonini2019,antonini2020,shi2020}. Using the initial stellar cluster mass distribution from \citet{Schaerer2011} in our population synthesis approach, we found that most of the stellar clusters that aggregate to form the NSC have initial masses between $\sim 10^{5.5}$ to $\sim 10^{6.5} \ \msun$. Only for the least sampled, most massive stellar clusters (see right end of Fig. \ref{fig:cluster-mass-distribution}), an IMBH with a mass of up to  $10^{5} \ \msun$ can form. Our results show that such stellar clusters are only involved in the assembly of the most massive NSCs. The final SMBH masses that we consider in this work are significantly influenced by growth in the NSC: this can be seen in the differences between Figs. \ref{fig:nsc=mass-vs-bhmass-no-growth} and \ref{fig:nsc=mass-vs-bhmass-growth300myr}. Therefore, changes to the IMBH masses that we sample would not significantly affect our predicted results in Fig. \ref{fig:nsc=mass-vs-bhmass-growth300myr} which are consistent with observed SMBH masses.

\subsubsection{GW recoil kick and escape velocities of NSCs}\label{subsec:gw-recoil-kick}

Another strong assumption in our work is that only merging IMBHs in the galactic nuclei with mass ratios less than 0.15 will be retained. The median GW recoil kicks for such mergers is expected to be less than a few hundred $\kms$ which is consistent with the escape velocities of denser and more massive NSCs \citep{fragione2020}. It is possible that the GW recoil kicks are sufficient to eject the merged BH in many NSCs \citep{gerosa2019,antonini2019,mapelli2021}. Our results show that two or more IMBHs are predominantly delivered in NSCs more massive than $10^{7} \ \msun$ and these NSCs will probably have higher escape velocities. More importantly, our calculations of the GW recoil kick magnitude for merging IMBH binaries assume a spin distribution for BHs which peaks close to 0.7 \citep{Lousto2012}. If BHs had lower spin values, these recoil kick velocities would be significantly smaller and then it may also be possible to retain the merged BHs even when the mass ratio of the merging BHs is larger than 0.15. If seed BHs grow in the galactic nuclei through episodic chaotic accretion of surrounding gas, it may be possible to maintain low spin values \citep{king2006,king2008}. It has also been suggested that IMBHs forming or growing through repeated mergers with other BHs will eventually have a decreased spin value. \citet{fragione2021} find that IMBHs ($\gtrsim 1000 \ \msun$) forming in dense massive clusters have spin values of $\sim 0.3$. Such spin magnitudes would result in a lower mean GW recoil kick value and may increase the retention probability of a seed SMBH without the need for significant BH growth through gas accretion.

\subsubsection{Prompt IMBH binary mergers and few-body interactions}\label{subsec:imbh-merger-interactions}

In our population synthesis approach, we assume that pairs of IMBHs will promptly merge within the NSC. We found in \citetalias{askar2021} that the GW merger time for a binary IMBH depends on the evolution of its semi-major axis and orbital eccentricity. The IMBH binary hardens and becomes more eccentric due to interactions with surrounding stars leading to GW mergers within a few $\sim$ 100 to 1000 Myr from the time of the IMBH binary formation \citep{rasskazov2019,SeddaAMB2019}. For NSCs where only two or fewer IMBH are delivered, we do not expect the prompt merger assumption to significantly change the results for final BH mass and occupation fraction. In few of the cases, particularly for high-mass NSCs, when multiple IMBHs are delivered to the NSC, we may have cases where few-body interactions between IMBHs can take place. In \citetalias{askar2021}, we found that for runs with three IMBHs in them, a binary-single encounter between the IMBHs can result in the third IMBH being scattered away. Such interactions can drastically influence the values of the orbital parameters of the binary IMBH and their subsequent evolution. In our population synthesis study, the delivery of IMBHs to the NSC is delayed due to the distribution of the delivery times of the merging stellar clusters. We expect that in more than 60 per cent of NSCs where more than two IMBHs are delivered, the delivery time of the third IMBH will be longer than few hundred Myr which may be sufficient time for the binary to merge due to GW radiation. It is also possible for such interactions to result in eccentricity fluctuations of the IMBH binary and they may be driven up to higher values where they may merge due to GW radiation \citep{hoffman2007,sobolenko2021}. 

\subsubsection{Other possible SMBH seeding mechanisms}

Lastly, it is also important to consider other possible massive BH seeding mechanisms that are fundamentally similar to the scenario proposed in this paper. An SMBH seed may still form in a dense and massive NSC even if none of the stellar clusters that merged to form the NSC contained an IMBH. The same processes that can lead to the formation of IMBHs within stellar clusters may occur in an NSC leading to the eventual formation of an SMBH seed. These processes include mergers of stars \citep{SPZ02,SPZ04,Giersz15,das2021} or stellar-mass BHs \citep{Miller02,rizutto2020,das2021,sedda2021,fragione2021}, tidal encounters \citep{stone2017,sakurai2019} and binary evolution of massive stars \citep{dicarlo2021,gonzalez2021,mapelli2021}. The scenarios involving massive stars would require the presence of young newly formed stars in the NSC. Additionally, growth of stellar-mass BHs and low-mass IMBH from the tidal capture and disruption of stars is most likely to occur in NSCs with central densities larger than few $10^6 \ \msun$ \citep{stone2017,tanikawa2021}. For dense stellar systems with high central velocity
dispersion values (larger than $50 \ \kms$), binary-single interactions will be unable to heat the cluster and prevent core collapse. This is because at a high velocity dispersion, binaries are sufficiently tight such that binary-single encounters will lead to stellar mergers and the formation of a massive central BH \citep{Miller2012}. Additionally, significant gas accretion by stellar-mass BHs in an NSC may lead to their growth and the eventual formation of an SMBH seed \citep{vesperini2010,Davies2011,leigh2014,natarajan2021,das2021}. Recently, \citet{kroupa2020} have suggested that it may be possible to form massive SMBH seeds within a few hundred Myr in dense gas-rich hypermassive starburst clusters of young spheroidal galaxies through the runaway growth of stellar-mass BHs. Gas inflow into these massive clusters would cause them to shrink, this will drive up their central velocity dispersion leading to the formation of a massive central BH \citep{Davies2011,kroupa2020} on short timescales. This process may explain the origin of high-redshift quasars. All these processes can lead to the in-situ formation and growth of an SMBH seed in an NSC. However, these processes are highly stochastic and will depend on the properties and the exact formation mechanism of the NSC during the early evolution of a galaxy. Other possible SMBH seeding mechanisms involve direct collapse of a significant amount of gas into massive BH seeds or through growth of BHs forming from population III stars \citep[for more details, please see][]{latif2016,greene2019rev,inayoshi2020,sassano2021}.






\subsection{Comparison with \citet{chassonnery2021}}\label{subsec:comparison-cd2021}

In a relevant recent work, \cite{chassonnery2021} considered the idea that that a number of massive
stellar clusters containing one or few IMBHs will end up in the nuclei of their host galaxy due to dynamical friction. They then studied the dynamical evolution of an extremely dense cluster of 400 IMBHs (each IMBH having a mass of $10^{4} \ \msun$) with direct \textit{N}-body simulations (including 2.5 post-Newtonian terms). They found that repeated mergers between binary IMBHs due to GW emission leads to the formation of a seed SMBH that ends up aggregating 20 per cent of the initial cluster mass. This seeding mechanism can lead to the rapid formation of a heavy seed SMBH in the centre of a dense cluster of IMBHs. However, the formation of such a dense cluster of IMBHs would require the aggregation of between several hundred and a few thousand massive stellar clusters in a galactic nucleus.

In \citetalias{askar2021}, we considered the simultaneous merger of three stellar clusters to form an NSC, where some of these clusters contain a single IMBH at their centre. Using \textit{N}-body simulations, we showed that in runs containing two or more IMBHs, an IMBH binary is likely to form at the centre of the merged cluster. The IMBH binary hardens and becomes more eccentric as it interacts with surrounding stars. We found that in some cases, the IMBH binary eccentricity acquires large values and GW emission can lead to the efficient merger of the IMBHs resulting in the formation of a seed SMBH that maybe retained in the NSC. In this scenario, only a few IMBHs are needed to be delivered to a galactic nucleus in order to seed the SMBH.

Despite the differences in the initial setup, both \citetalias{askar2021} and \citet{chassonnery2021} results suggest that IMBH binary mergers in a galactic nucleus may play an important role in seeding SMBHs. The two studies differ in their results for the calculation of GW recoil kicks (compare Fig. 16 in \citetalias{askar2021} with Fig. 3 in \citet{chassonnery2021}). This difference can be attributed to the assumed spin distributions of the merging BHs (see Section \ref{subsec:gw-recoil-kick}). While \citetalias{askar2021} draw their BH spin distribution from the results of \citet{Lousto2012}, \citet{chassonnery2021} draw the BH spins uniformly between zero and one. The latter approach results in significantly lower GW recoil kick magnitudes and therefore \citet{chassonnery2021} find that the retention probability of a merged BH is higher. Additionally, since the IMBH cluster in \citet{chassonnery2021} is extremely dense, its escape speeds are of the order of $10^{3} \kms$ which makes it easier to retain merged BHs following GW recoil kicks. 

\section{Conclusions}\label{sec:conclusion}

In this work, we explored the idea that an NSC forms before the SMBH through the merger of several stellar clusters that may contain IMBHs. In \citetalias{askar2021}, we showed that these IMBHs can subsequently merge and grow in the NSC to form an SMBH seed. To check the observable consequences of this proposed SMBH seeding mechanism, we carried out a population synthesis study by creating a synthetic population of galaxies and their NSCs using observed distributions and scaling relations. For each galaxy, we randomly drew and aggregated stellar cluster masses until we could reproduce the NSC mass. Using results from previous works on IMBH formation in stellar clusters \citep{askar2017,morawski2018,Sedda2019}, each of these aggregated stellar clusters was assigned a probability to bring along an IMBH. Through this population synthesis approach, we were able to determine the likelihood of having seed SMBHs delivered in an NSC and obtain the occupation fraction of massive BHs that arise from this seeding mechanism as a function of NSC mass (see Fig. \ref{fig:occupation-frac-nsc-mass}) and galaxy stellar mass (see Figs. \ref{fig:bh-occupation-frac-galaxy-stellar mass} and \ref{fig:massive-bh-occupation-frac-galaxy-stellar mass}). Our results show that through this seeding mechanism, the occupation fraction of seed SMBHs increases with NSC and galaxy stellar mass. Without considering BH growth, we find that GW recoil kicks can eject merged BHs from the NSC and that the occupation fraction of a seed BH in an NSC is about 0.6 for galaxies with stellar masses larger than about $10^{10} \ \msun$. If the BH is allowed to grow inside the NSC than the mass ratio of the merging BHs can decrease significantly, resulting in lower GW recoil kick values and a higher likelihood of retaining the merged BH in the NSC. In this case, the occupation fraction of BHs in more massive galaxies can be as high as up to 80 per cent. These results are consistent with occupation fraction results from X-ray and dynamical studies of galactic nuclei in the local Universe \citep{miller2015,Nguyen2019,greene2019rev}.

For galaxies where there are measurements or constraints on NSC mass and SMBH mass, we categorized them into three different types in Fig. \ref{fig:obs-nsc-vs-bh}. Our main findings indicate that less massive elliptical and spiral galaxies (``A'' type galaxies) with NSC masses less than about few $10^{6} \ \msun$ have at the most 50 per cent probability of containing a single SMBH seed.  Therefore, our proposed seeding mechanism can explain the absence of an SMBH in relatively low-mass galaxies like M33. The dry merger of only a few stellar clusters is needed to obtain the NSC mass in these galaxies and therefore the likelihood of any of those clusters delivering an IMBH to the NSC is lower.

For more massive galaxies with NSC masses larger than few $10^{6} \ \msun$ (``B'' type galaxies), the likelihood of having an SMBH seed increases to about 60-80 per cent. More stellar clusters need to be aggregated to obtain the NSC mass for the galaxies, therefore, the likelihood of at least one of those clusters delivering an IMBH to the NSC is higher. The presence of gas in the nuclei of these galaxies may lead to BH growth through accretion and may also contribute to the growth of the NSC through in-situ star formation. Our population synthesis model reproduces the diversity and correlations seen in the $\rm M_{\rm NSC} - M_{\rm BH}$ plane. Allowing for BHs to grow at an average mass accretion rate that is 10 per cent of the Eddington rate also allows us to recover observed SMBH masses. BH and NSC growth through gas may also explain the differences in galaxy stellar mass, BH mass and NSC mass between ``A'' and ``B'' type galaxies. As illustrated in Fig. \ref{fig:illustration}, it is likely that the gas fraction in ``A'' type galaxies was lower than that in the ``B'' type galaxies.

More massive elliptical galaxies (``C'' type galaxies) have massive SMBHs but low-mass NSCs. It is likely that these galaxies formed from the major merger of spiral galaxies. Such mergers may result in these galaxies having more massive SMBH. The merging of galaxies and the presence of an SMBH may lead to the destruction of the NSC and quenching of its growth. The processes by which this can occur were discussed in Section \ref{subsec:c-type}.

We presented a detailed discussion of the various uncertainties and assumptions that go into our population synthesis work in Section \ref{subsec:caveats}. In a subsequent paper, we will examine how this proposed seeding mechanism could be verified and constrained with GW observations of extreme and intermediate-mass ratio inspirals \citep{as2007,fragione2018d,sedda2020aschen,raveh2021} involving IMBHs \citep{matsubayashi2004,amaro2006,fregeau2006} through planned future missions such as Laser Interferometer Space Antennae \citep[LISA;][]{lisa2017}.
\vspace{-0.5cm}
\section*{Acknowledgements}

We would like to thank the reviewer for their comments that helped in improving the quality of the manuscript. AA would like to thank Ammar Askar, Swayamtrupta Panda, Alister Graham, Alessandra Mastrobuono Battisti, Pavel Kroupa and Fabio Pacucci for useful discussions. AA and RC acknowledge support from the Swedish Research Council through the grant 2017-04217. AA also acknowledges support from the Royal Physiographic Society of Lund and the Walter Gyllenberg Foundation for the research grant: `Evolution of Binaries containing Massive Stars'. AA was also supported by the Carl Tryggers Foundation for Scientific Research through the grant CTS 17:113 (2018-2020). Some of the figures in the paper were made using the Seaborn data visualization library for Python.
\vspace{-0.5cm}
\section*{Data Availability Statement}

The code written to carry out the population synthesis (see Section \ref{sec:sec2-mock} and Fig. \ref{fig:flow-chart}) in this work and the data generated from it will be shared on request to the corresponding author. 

\bibliographystyle{mnras}
\bibliography{biblio} %



\appendix
\section{Heatmaps for IMBH Delivery and Retention}\label{appendixA}
\begin{figure*}
	  \includegraphics[width=0.49\linewidth,scale=13]{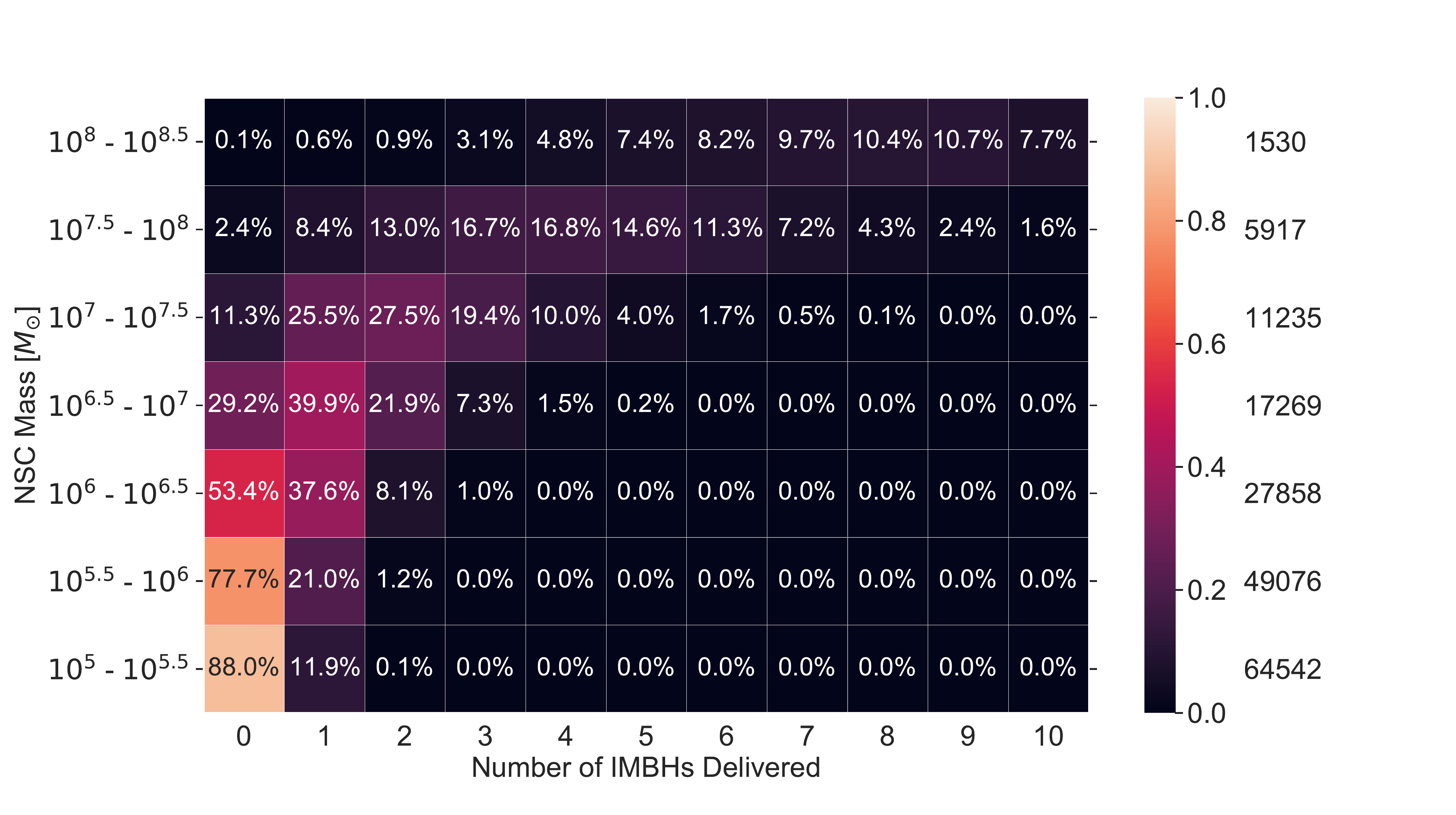}
    \includegraphics[width=0.49\linewidth,scale=13]{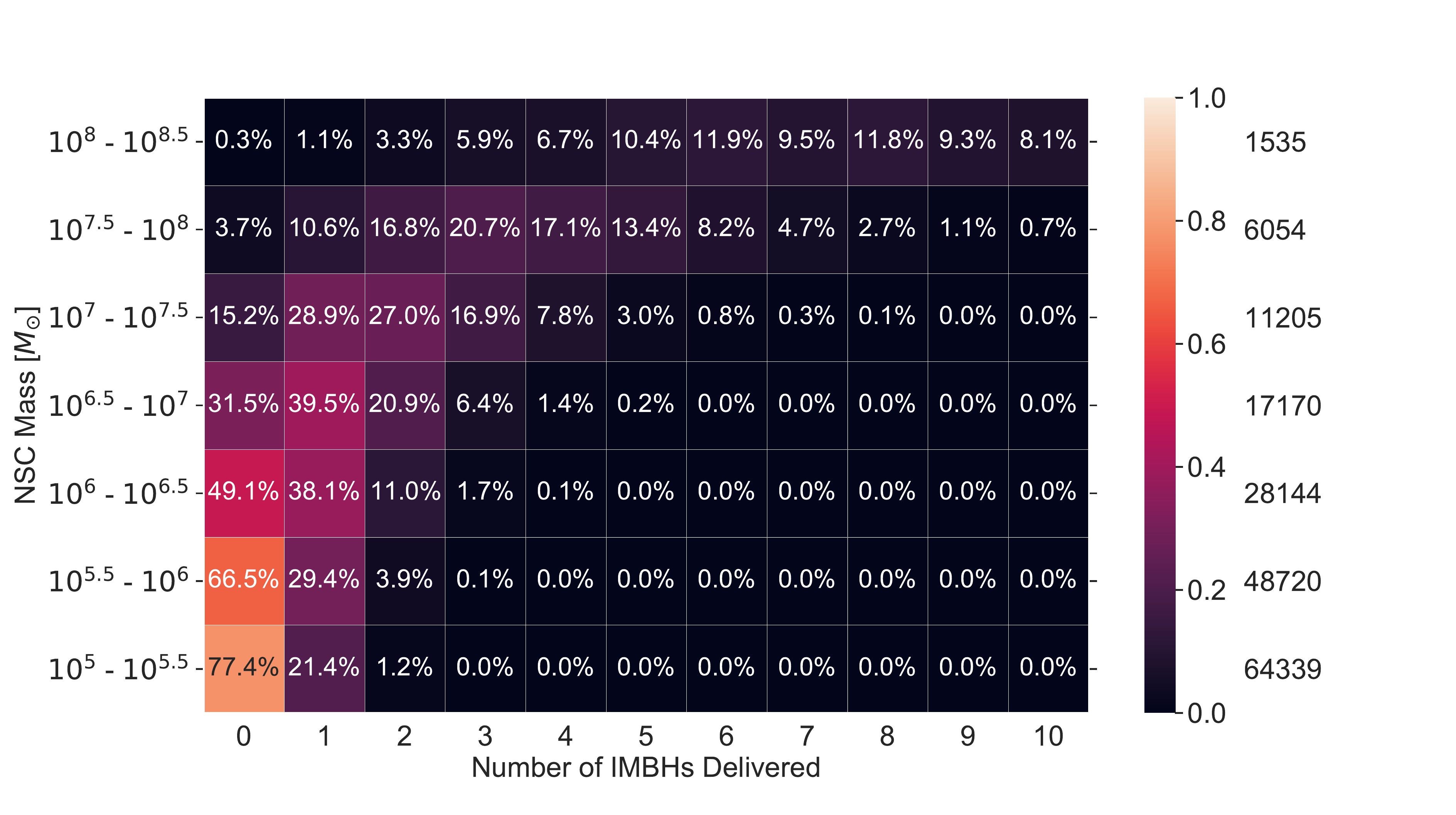}
	\caption{The percentage of delivered IMBHs that correspond to different mass bins of NSC mass on the y-axis in increasing order. For the left heatmap, the high natal kick assumption for stellar-mass BHs is used when considering the probability that a stellar cluster forms an IMBH (see Table \ref{tab:reaslitic}) . For the right heat map, the low kick natal kick for stellar-mass BH assumption is used \citep{belczynski2002}. The numbers towards the right of the colour bar indicate the number of NSCs in each of the bins.}
    \label{fig:heatmap-all}
\end{figure*}

\begin{figure*}
	  \includegraphics[width=0.49\linewidth,scale=9.0]{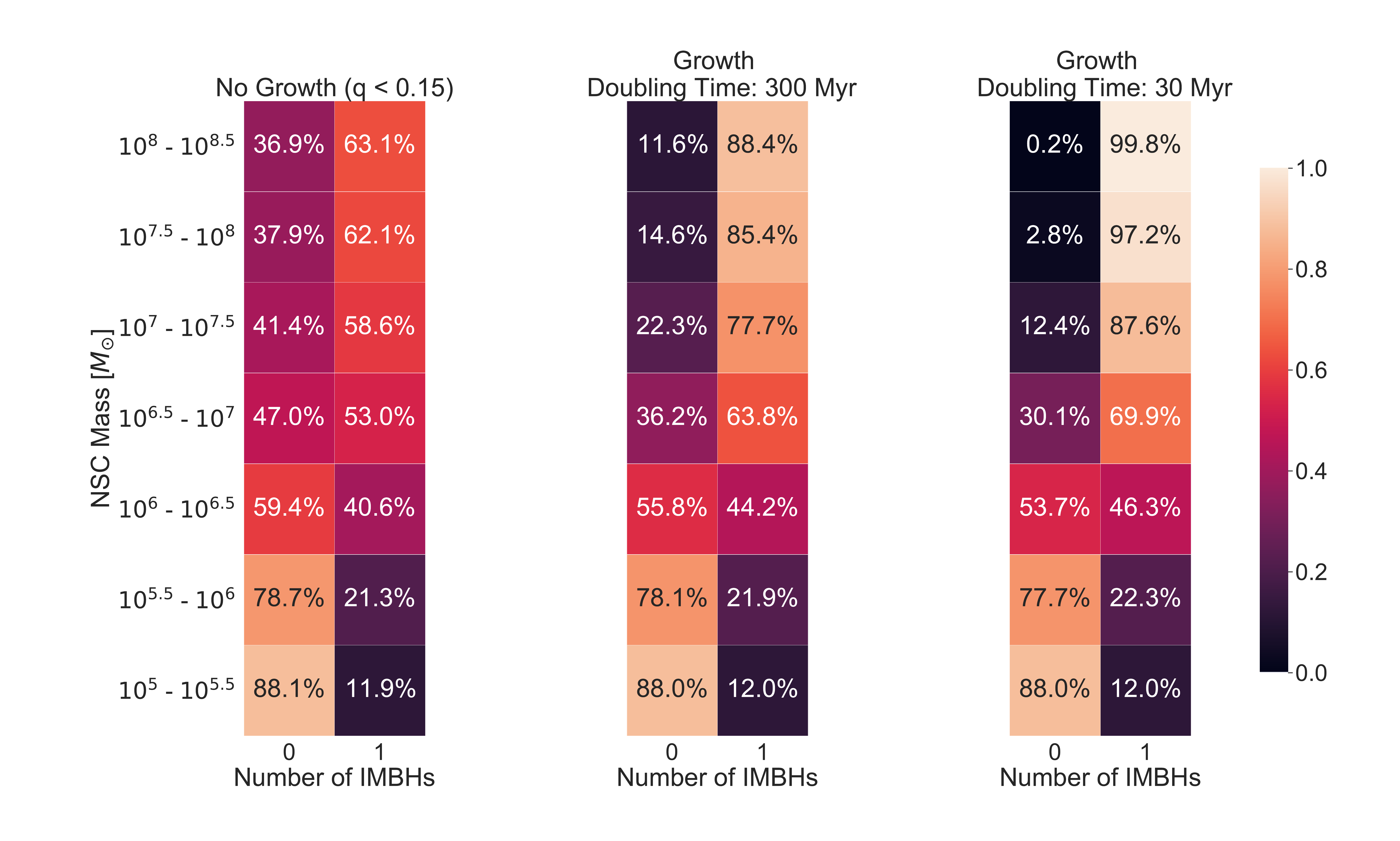}
	  \includegraphics[width=0.49\linewidth,scale=9.0]{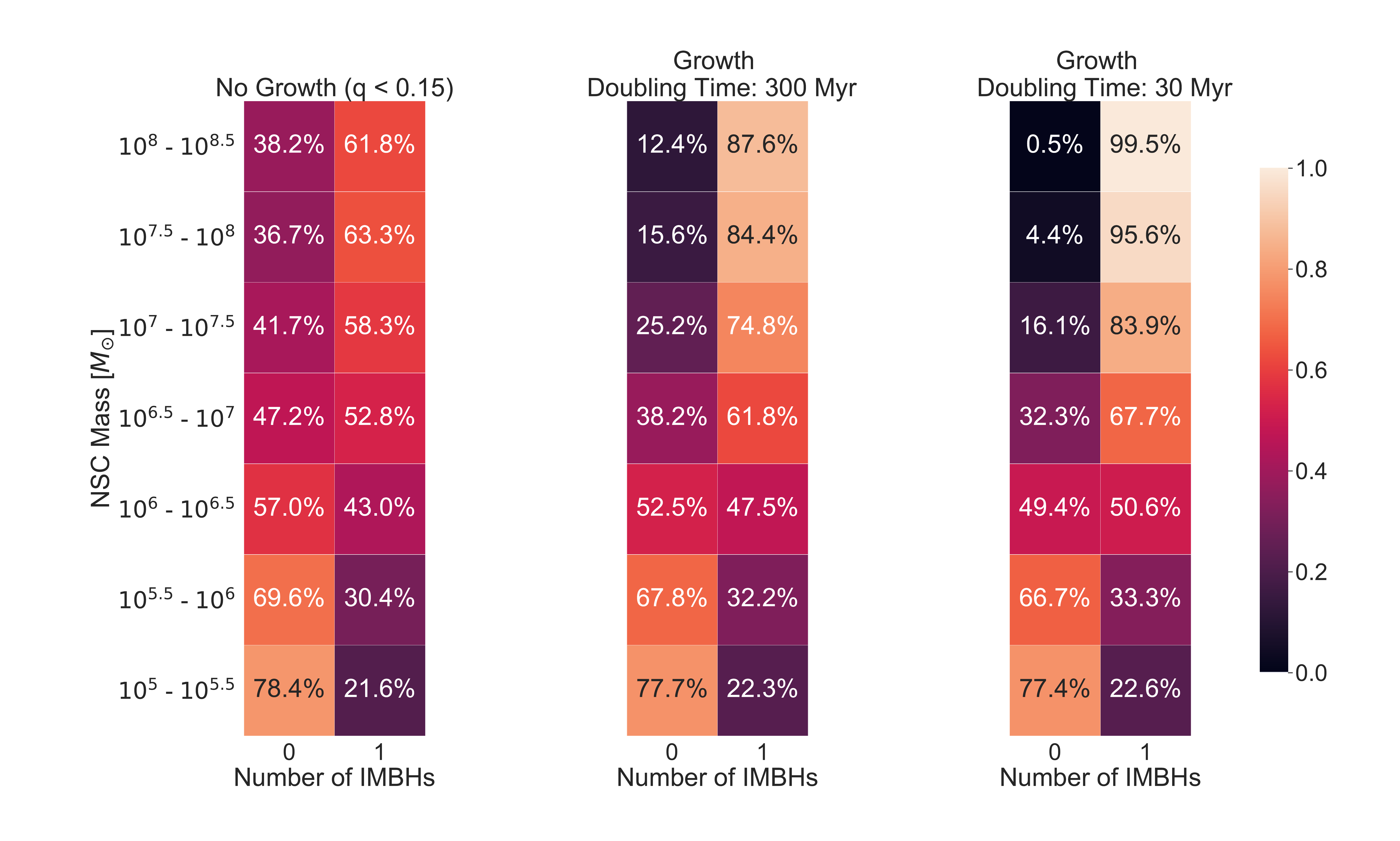}
	\caption{Retention of an SMBH seed for different assumptions about BH growth. The three left panels show the results for the high natal kick assumption and the right three panels show the results for the low natal kick assumptions. Each panel considers different cases for IMBH retention. For the no growth panel, if two IMBHs merge then the product is only retained if the mass ratio of the merging BHs is less than 0.15. For the two remaining panels, we allow the BHs to grow before the merger of the BHs occurs}
    \label{fig:heatmap-processed}
\end{figure*}

In Fig. \ref{fig:heatmap-all}, we present an alternative illustration of the percentage of our NSCs models in which a particular number of IMBH are delivered (see Fig. \ref{fig:histo-delivered-low}). Each grid in the heatmap  show the percentage of NSCs (binned in different mass ranges indicated on the vertical axis) in which the total number of delivered IMBH is given on the horizontal axis. The left panel in Fig. \ref{fig:heatmap-all} corresponds to the case where we draw the probabilities for stellar clusters that form the NSC to contain an IMBH based on the high natal kick assumption for stellar-mass BH formation in stellar cluster simulation models (see Table \ref{tab:reaslitic}). There are only small differences in the number of delivered IMBHs to the NSC between the high and low natal kicks assumptions (left and right panel in Fig. \ref{fig:heatmap-all}). The probability of IMBH formation in the low kick assumption is higher for GC with masses less than $5 \times 10^{5} \ \msun$, this leads to a slightly higher number of NSCs in which at least one IMBH has been delivered. The SMBH seeding mechanism proposed here naturally accounts for a lower occupation fraction of SMBHs in galaxies with low-mass NSCs as compared to galaxies with higher mass NSCs. This is discussed further in Section \ref{sec:discussion}.

The heatmaps in Fig. \ref{fig:heatmap-processed} shows the percentage of NSCs in which either zero or one IMBH remains after considering IMBH mergers (similar to Fig. \ref{fig:histo-processed}) for three different cases:
\begin{enumerate}
    \item \textit{No BH Growth} (see first column in Fig. \ref{fig:heatmap-processed})
    \item \textit{BH Growth with mass doubling time of 300 Myr} (second column in Fig. \ref{fig:heatmap-processed})
    \item \textit{BH Growth at Eddington rate} (third column in Fig. \ref{fig:heatmap-processed})
\end{enumerate}



\bsp	
\label{lastpage}
\end{document}